\documentclass[runningheads]{llncs}
\usepackage{graphicx}
\usepackage{soul}
\setstcolor{red}
\usepackage[dvipsnames]{xcolor}
\usepackage[ruled]{algorithm2e} 

\SetAlFnt{\small}
\SetAlCapFnt{\small}
\SetAlCapNameFnt{\small}
\SetAlCapHSkip{0pt}
\IncMargin{-\parindent}
\usepackage{amsfonts}
\usepackage{diagbox}
\usepackage{amsmath}
\usepackage{mathtools}
\usepackage{wrapfig}
\usepackage{booktabs}
\renewcommand{\Re}{ \mathbb{R} }
\newcommand{\Z}{ \mathbb{Z} }

\newcommand{\faces}{ \mathcal{F} }

\newcommand{\vertices}{ \mathcal{V} }

\usepackage{color} 
\definecolor{color1}{rgb}{0.5,0.1,0.1}
\definecolor{color2}{rgb}{0.1,0.5,0.1}
\definecolor{color3}{rgb}{0.1,0.1,0.5}
\definecolor{color4}{rgb}{0.5,0.1,0.5}
\definecolor{color5}{rgb}{0.5,0.5,0.1}

\usepackage{caption}
\usepackage{subcaption}
\usepackage{multirow}
\renewcommand{\vec}[1]{ \ensuremath{\mathbf{#1} } }
\begin{document}
\title{A direct geometry processing cartilage generation method using segmented bone models from datasets with poor cartilage visibility}
\titlerunning{Cartilage generation method}
%
 \author{Faezeh Moshfeghifar \inst{1} \and Max Kragballe Nielsen \inst{1}\and José D. Tascón-Vidarte \inst{1} \and
Sune Darkner\inst{1} \and Kenny Erleben\inst{1}}
%
%
\authorrunning{F. Moshfeghifar et al.}
%
\institute{
Department of Computer Science, University of Copenhagen, Denmark.\\
\email{\{famo,max,jota,darkner,erleben\}@di.ku.dk}
}
\maketitle              
\begin{abstract}
We present a method to generate subject-specific cartilage for the hip joint. Given bone geometry, our approach is agnostic to image modality, creates conforming interfaces, and is well suited for finite element analysis. We demonstrate our method on ten hip joints showing anatomical shape consistency and well-behaved stress patterns. Our method is fast and may assist in large-scale biomechanical population studies of the hip joint when manual segmentation or training data is not feasible.

\keywords{Cartilage Generation \and Hip Joint \and Finite Element Analysis.}
\end{abstract}

\section{Introduction}
The femur and pelvic bones come together in the hip joint (HJ), where the femoral head articulates within the acetabulum's lunate surface. The lunate surface and all of the femoral head, except the fovea pit, are covered by cartilage tissue allowing unhindered motion in the HJ~\cite{standring2020gray}.
The stress distribution on these articular surfaces is often analyzed either experimentally~\cite{c8-1,c8-2}, using finite element (FE) analysis~\cite{c3,c7,jorge2014finite}, or discrete element analysis~\cite{c6,c8,c11}. The analysis results provide valuable information for studying both healthy and dysplastic hips~\cite{c1,c6,c7,c8,c10}. The accuracy of these results is highly dependant on the HJ geometries.  

HJ morphology is segmented directly from computed tomography (CT)~\cite{c2,c3,c6,c7,c10} and magnetic resonance imaging (MRI) images~\cite{c12}. CT and MRI scans contain high contrast images of the bones with clearly visible interfaces. However, it is difficult to identify cartilage tissue due to the tight space in HJ~\cite{c16} or low image resolution~\cite{R13}. CT arthrography and manual traction enhance the cartilage visibility and increase the joint space at the cost of being an invasive intervention~\cite{c2,c7}. 

This work presents a method to generate subject-specific cartilage models where cartilage segmentation is not feasible. Our method uses only the bone geometries coming together at the HJ and reconstructs the articulating cartilage surfaces based on bone curvature measures and distance measures.
Our solution is fast, is independent of image modality, and preserves anatomical properties compared to previous methods.
We validated our method by comparing the anatomical properties to other works, as detailed in section~\ref{sec:results}~\cite{anderson2008validation,c15,lazik2016,schmaranzer2019automatic}. Additionally we tested their performance in a FE analysis setup. Our goal is to obtain consistent and stable cartilage models for FE analyses. For now, patients with HJ atypical geometries or pathology are out of the scope of this method.

\section{Related work}
The femur and pelvis can be segmented using manual~\cite{c3,c7}, semi-automatic~\cite{c2,c6,lopes2020hip}, or fully automated approaches~\cite{c9,c10,c12}.

When cartilage is manually segmented, the thickness and interfaces are either delineated from the images, assigned uniformly, or approximated by a radius representing the smoothed joint space mid-line~\cite{c3,c6,c7}. This manual process is time-consuming and is limited by the users' clinical expertise and image resolution. Our method ensures conforming surfaces and computes thickness directly from bone locality without any need for clinical expertise.

In semi-automated segmentation approaches, the initial cartilage geometry is delineated automatically, starting from user-defined landmarks. Additional refinement is needed to eliminate rough surfaces, holes, and irrelevant connected tissues~\cite{c2,c6}. In contrast, our method only relies on bone surfaces and does not need landmarks or refinement.

With the advancements in computational morphometrics, automated cartilage segmentation is possible by combining bone statistical shape models with population-averaged cartilage thickness maps~\cite{c11} or geometric constraints~\cite{c12}. Automatic segmentation is a cost-effective approach and reduces the need for pre-processing data~\cite{c11}. However, segmented data and prior information is needed to train such models. In contrast, our method does not require training data and is agnostic to data-shift problems.

Other approaches simplify the joint to a perfect ball in socket joint and fit the bone-cartilage and cartilage-cartilage interfaces to spheres or rotational conchoids~\cite{c3,c5,c8}. These simplifications can have an effect on the cartilage contact pressure and contact area~\cite{c3}. Our work provides non-uniform cartilage thickness while maintaining conforming interfaces.

\section{Methodology} \label{sec:method}
The proposed cartilage generation method is based only on bone geometry and distance measures. Thus, the algorithm is sensitive towards the quality of the segmented bone mesh, which we assume has been preprocessed to remove irregularities and segmentation artifacts. 
We will focus on the hip joint and generate cartilage with respect to either the pelvis or femur, depending on which bone the cartilage attaches to. We will need to generate cartilage twice: once for the femur and once for the pelvis. We define the geometry of each bone as the vertices and faces ($\vertices, \faces$) of a triangle mesh, where $\vertices \in \Re^{N \times 3}$ and $\faces \in \Z_{0+}^{K \times 3}$ for a mesh consisting of $N$ vertices and $K$ faces. We refer to the bone where cartilage attaches as the \textit{primary bone}, $(\vertices_{P}, \faces_{P})$, and the other bone as the \textit{secondary bone}$, (\vertices_{S}, \faces_{S})$.
The generated cartilage will be a triangle surface mesh $(\vertices_{C}, \faces_{C})$. Our method can be summarized as the steps:
\begin{description}
    \item[Distance Filtering:] Select an initial subset of the primary bone as the \textit{bone-attached} cartilage region based on the distance to the secondary bone;
    \item[Curvature-Based Region Filling:] Apply our curvature-based region filling approach to ensure the bone-attached cartilage region extends to anatomical lines;
    \item[Extrusion:] Extrude a subset of the bone-attached region towards the secondary bone;
    \item[Harmonic Boundary Blending:] Interpolate between the boundary of the extruded and bone-attached regions to create a soft blend.
\end{description}

\subsection{Distance Filtering} \label{sec:distance:filtering}
\begin{figure}[h!]
    \centering
    \begin{subfigure}[t]{0.2\textwidth}
        \centering
        \includegraphics[width=\textwidth]{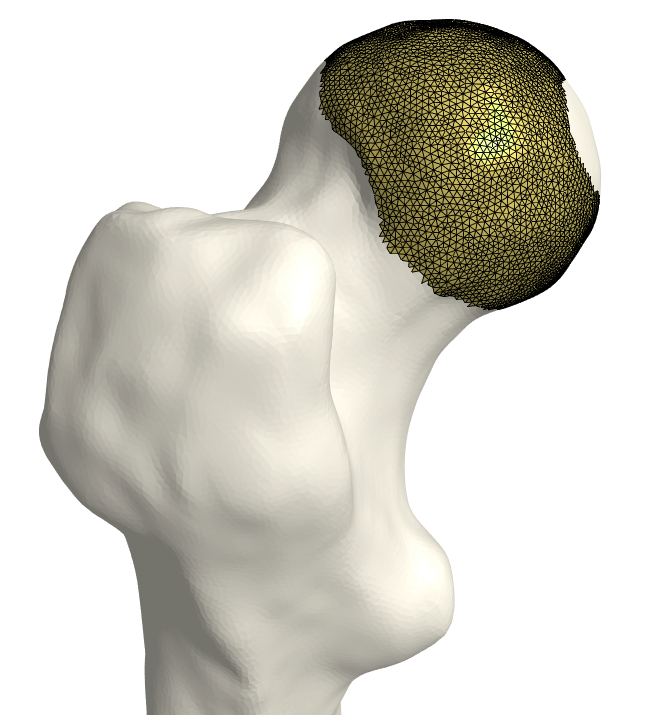}
        \caption{Initial.}
        \label{fig:initial:guess:femur:no:trimming}
    \end{subfigure}
    \hspace{1cm}
    \begin{subfigure}[t]{0.2\textwidth}
        \centering
        \includegraphics[width=\textwidth]{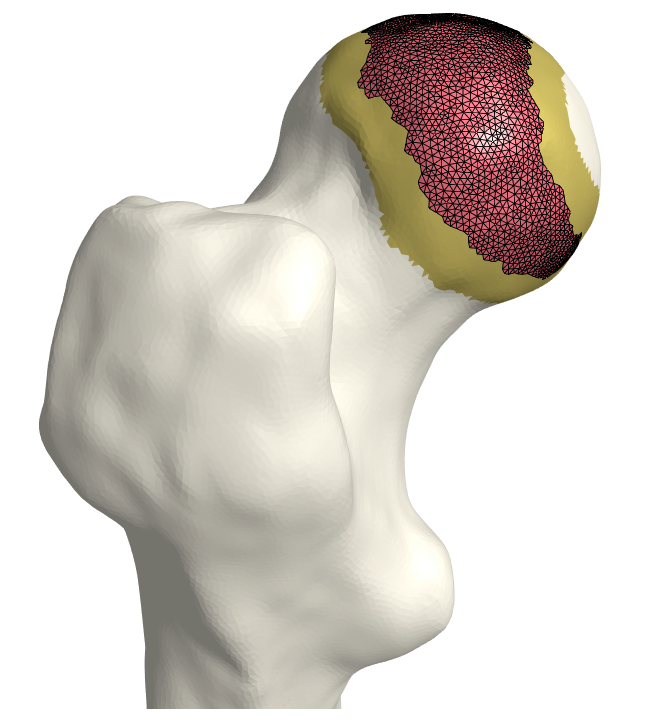}
        \caption{Trimming.}
        \label{fig:initial:guess:femur:trimming}
    \end{subfigure}
    \hspace{1cm}
    \begin{subfigure}[t]{0.2\textwidth}
        \centering
        \includegraphics[width=\textwidth]{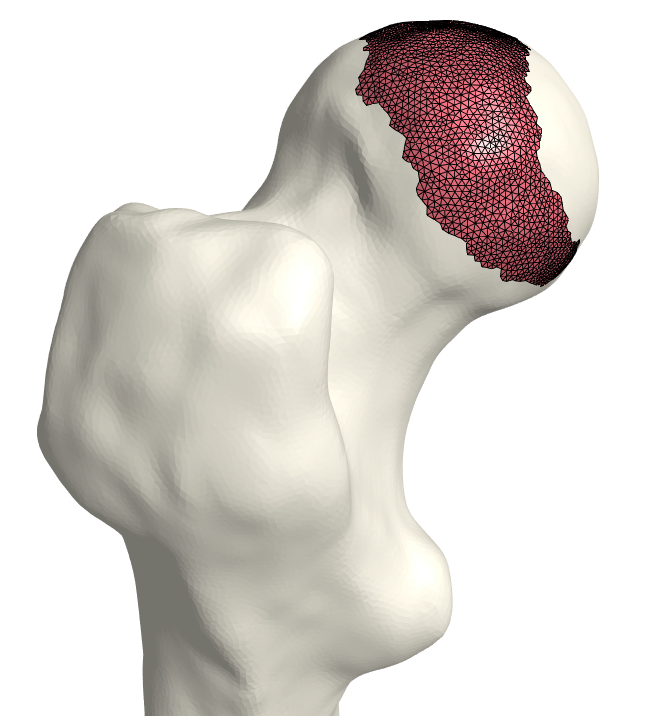}
        \caption{Max. cluster.}
        \label{fig:initial:guess:femur:trimming:and:single}
    \end{subfigure}
        \begin{subfigure}[b]{0.2\textwidth}
        \centering
        \includegraphics[width=\textwidth]{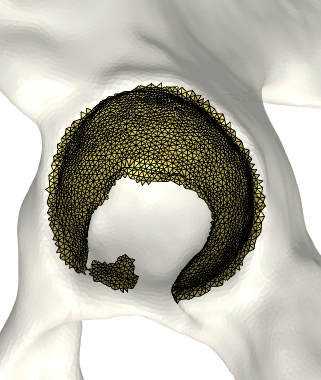}
        \caption{Initial.}
        \label{fig:initial:guess:pelvis:no:trimming}
    \end{subfigure}
    \hspace{1cm}
    \begin{subfigure}[b]{0.2\textwidth}
        \centering
        \includegraphics[width=\textwidth]{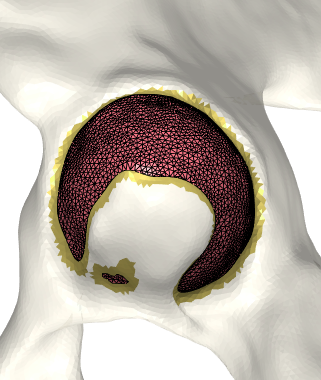}
        \caption{Trimming.}
        \label{fig:initial:guess:pelvis:trimming}
    \end{subfigure}
    \hspace{1cm}
    \begin{subfigure}[b]{0.2\textwidth}
        \centering
        \includegraphics[width=\textwidth]{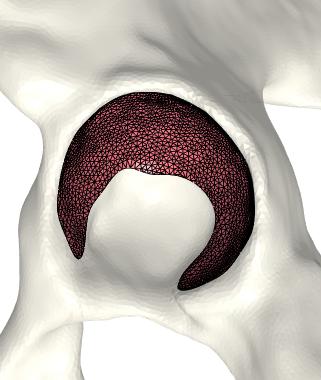}
        \caption{Max. cluster.}
        \label{fig:initial:guess:pelvis:trimming:and:single}
    \end{subfigure}
    \caption{Computing initial bone-cartilage interface for femur (\textbf{(a)-(c)}) and pelvis (\textbf{(d)-(f)}). Initially, in \textbf{(a)} and \textbf{(d)}, triangles on the surface of the femur and pelvis are selected based on the distance to the opposite bone (Eq.~\eqref{eq:distance:filtering}). Afterwards, in \textbf{(b)} and \textbf{(e)}, the initial sets are trimmed to remove triangles that only have one neighbour. The trimmed region can be seen in red on top of the initial guess in yellow. Finally, we select the biggest cluster of triangles as our initial estimate of the bone-attached region for both femur and pelvis (\textbf{(c)} and \textbf{(f)}).
    }
    \label{fig:initial:guess:femur}
\end{figure}

For now, we will focus on the femoral side of the hip joint and describe the femur as the primary bone and the pelvis as the secondary. We select faces on the primary bone, which will serve as an initial guess of the bone-attached cartilage region. We base our choice on the distance between the face barycenters of the primary bone and the secondary bone vertices. That is, provided the distance parameter, $\delta$, we construct the set of faces,
\begin{align}
    \{\vec f \in \faces_{P} : \min\limits_{\vec v \in \vertices_S} \lVert \mathbf{BC}(\vec f) - \vec v \rVert \leq \delta\}
\label{eq:distance:filtering}
\end{align}
where $\mathbf{BC}(\vec f)$ is the barycenter of face $\vec f$. We refer to this subset of faces and vertices as $\faces_{C}^{D}$ and its corresponding vertices as $\vertices_{C}^{D}$. The distance filter parameter, $\delta$, should be based on the gap between the femur and pelvis, as it determines the initial approximation quality. The femur's initial estimate must be located above the anatomical line of the femoral head (Fig.~\ref{fig:initial:guess:femur:no:trimming}).
Otherwise, we do not enforce any restrictions on $\delta$. To provide additional robustness to the initial guess, we trim the outer boundary by removing layers from its outer rim and discarding faces with two boundary edges.
The trimming helps ensure the initial estimate does not become too large, and in particular, makes it simpler to avoid selecting triangles crossing the natural ridges of the bone.

The pelvis is more sensitive to the distance filtering parameter than the femur, as most of the cartilage exists in a plateau in the pelvic socket.
This implies the bone-attached region does not require any additional refining, as the initial estimate accurately aligns with anatomical lines (Fig.~\ref{fig:initial:guess:pelvis:no:trimming}).
The distance filtering step typically results in fragmented bone-attached regions, and before we proceed, we discard all but the largest of these regions (Fig.~\ref{fig:initial:guess:pelvis:trimming:and:single}).
We can safely discard the smaller regions, as we grow the cartilage during the next step of our algorithm to incorporate all triangles within an area of similar curvature.


\subsection{Curvature-Based Region Filling}
\label{sec:curvature:based:region:filling}
\begin{figure}[h]
    \centering
    \begin{subfigure}[b]{0.215\linewidth}
        \centering
        \includegraphics[width=\linewidth]{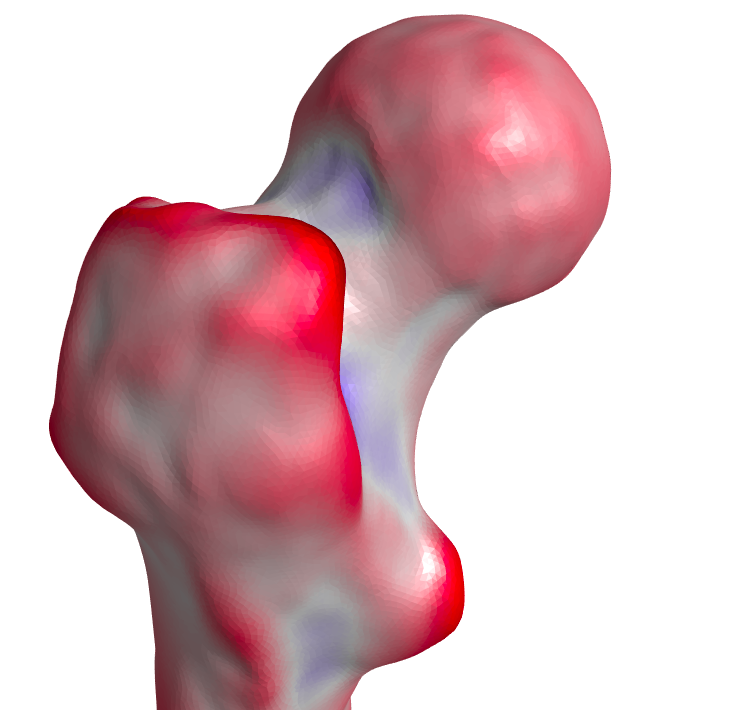}
        \caption{$\mathcal{N} = 10$.}
    \end{subfigure}
    \begin{subfigure}[b]{0.215\linewidth}
        \centering
        \includegraphics[width=\linewidth]{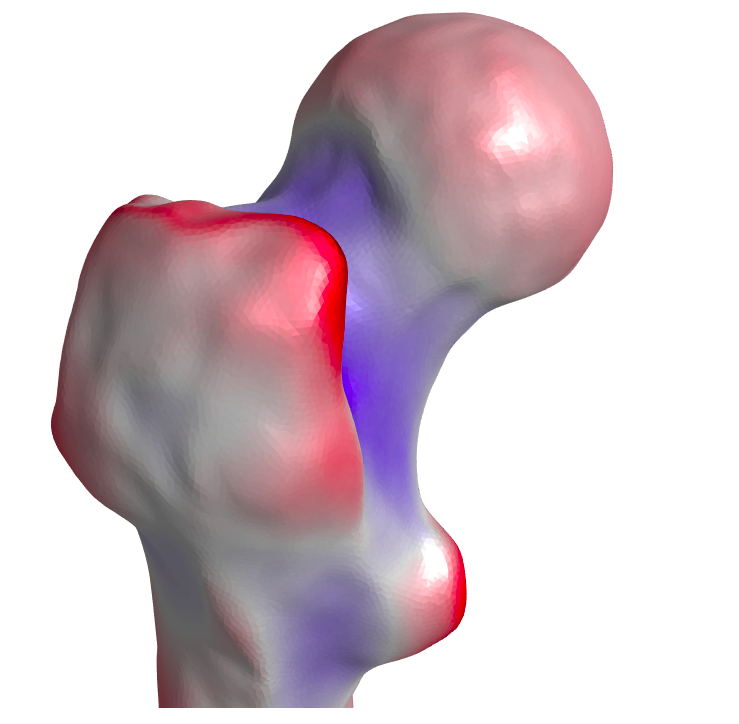}
        \caption{$\mathcal{N} = 20$.}
    \end{subfigure}
    \begin{subfigure}[b]{0.2\linewidth}
        \centering
        \includegraphics[width=\linewidth]{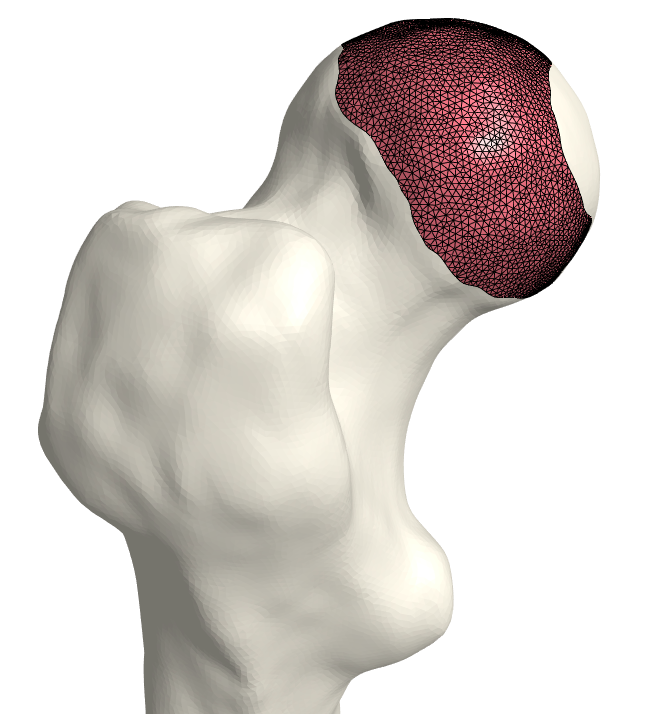}
        \caption{10 iterations.}
        \label{fig:grow:b}
    \end{subfigure}
    \begin{subfigure}[b]{0.2\linewidth}
        \centering
        \includegraphics[width=\linewidth]{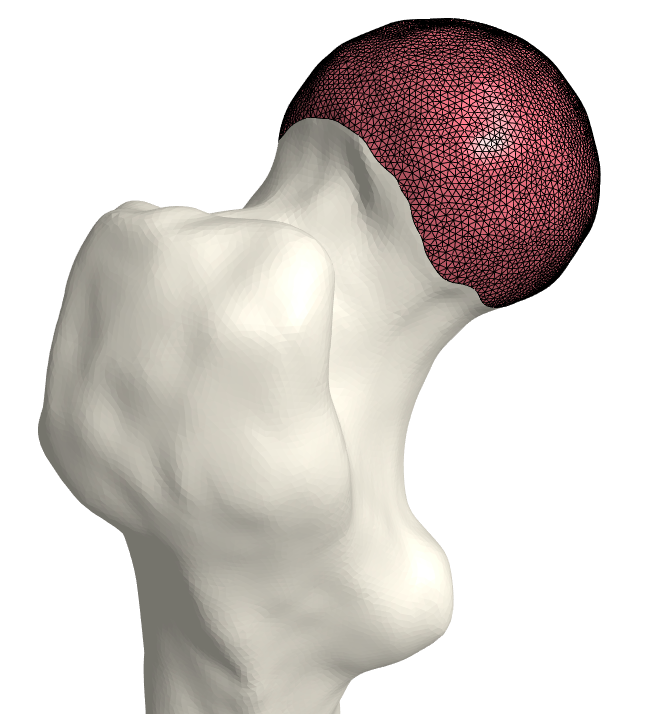}
        \caption{End result.}
        \label{fig:grow:c}
    \end{subfigure}
    \caption{\textbf{(a)-(b)} The effect of changing the neighborhood size, $\mathcal{N}$, in terms of mean curvature. Red regions correspond to positive curvature directions, while blue regions correspond to negative directions.
    Note that smaller variations in the surface geometry influence the curvature measure less as the neighborhood size increases.
    \textbf{(c)-(d)} Observe in Fig~\ref{fig:initial:guess:femur:trimming:and:single} the initial estimate of the bone-attached region, in \textbf{(c)} the bone-attached region after 10 iterations of Equation~\eqref{eq:grow}, and
    in \textbf{(d)} the final result, when no more faces satisfy Equation~\eqref{eq:grow}.}
    \label{fig:neighborhood:size}
\end{figure}

The initial estimate of the bone-side femoral cartilage does not yet fully cover the cartilage zone on the femoral head (Fig.~\ref{fig:initial:guess:femur:trimming:and:single}). The femoral cartilage border is observed as a change in the curvature between the femoral head and the neck.
In this work, the principal curvatures, $\kappa_{\min}$ and $\kappa_{\max}$, are computed by fitting local frames to a neighborhood around each vertex \cite{panozzo2010efficient}.
This method approximates the Laplace-Beltrami operator in the immediate neighbourhood of each vertex to save computation time.
The neighborhood size, $\mathcal{N}$, can be used as a scaling factor to control how many smaller variations to include in the curvature estimation (Fig~\ref{fig:neighborhood:size}), and when $\mathcal{N}$ is sufficiently large we will arrive at the same curvatures as can be computed using the mesh laplacian.

Denoting the boundary of $\faces_{C}^{D}0$ as $\Gamma_{C}^{D}$, we grow the cartilage such that,
\begin{align}
    \faces_{C}^{D} \leftarrow \faces_{C}^{D} \cup
    \big\{\vec f \in \Gamma_{C}^{D} : \kappa_{1} \leq \kappa_{\vec f} \leq \kappa_{2} \big\}
    \label{eq:grow}
\end{align}
where $\kappa_1$ and $\kappa_2$ denote the minimal and maximal curvature of the region at the femoral
head and $\kappa_{\vec f}$ is the curvature of face $\vec f$.
We grow the cartilage iteratively using Equation~\eqref{eq:grow} until the region contains all faces in $\faces_{C}^{D}$ (Fig.~\ref{fig:grow:c}).

\subsection{Extrusion}
\label{sec:extrusion}
As this is a trait of healthy cartilage, we expect the generated geometry to have a high degree of congruence.
We extrude vertices based on the minimal distances between the bones and choose the midpoint as the extrusion height to guarantee congruence. This approach ensures the conformity of the cartilage-cartilage interface. We want the shape to thin out as we get further away from the contact center to mimic the real cartilage shape. The faces to extrude, are chosen to be a copy of the bone-attached estimate, $(\vertices_{C}^{D}, \faces_{C}^{D})$, and are denoted by $(\vertices_{C}^{E}, \faces_{C}^{E})$: the extruded geometry. We assign height to each vertex in the extrusion region, $\vec v \in \vertices_{C}^{E}$, based on the smallest distance to the second bone as,
\begin{align}
    \vec v \leftarrow \vec v + \frac{1}{2} \vec n \min\limits_{\vec v_{S} \in \vertices_{S}} \lVert \vec v - \vec v_{S} \rVert_{2}
    \label{eq:extrusion}
\end{align}
where $\vec n$ is the unit outward normal direction of vertex $\vec v$. For the pelvis, the faces we extrude are a copy of the bone-attached region. As the pelvic bone-attached region does not need growing, we shrink the set of faces to be extruded by applying the same trimming approach as described in Section~\ref{sec:distance:filtering}. We then extrude the reduced set of faces (Equation~\eqref{eq:extrusion}), using the distances from the pelvis to the femur.

\subsection{Harmonic Boundary Blending}
\label{sec:harmonic:boundary:blending}
\begin{figure}
    \centering
    \begin{subfigure}[b]{0.25\linewidth}
     \centering
     \includegraphics[width=\linewidth]{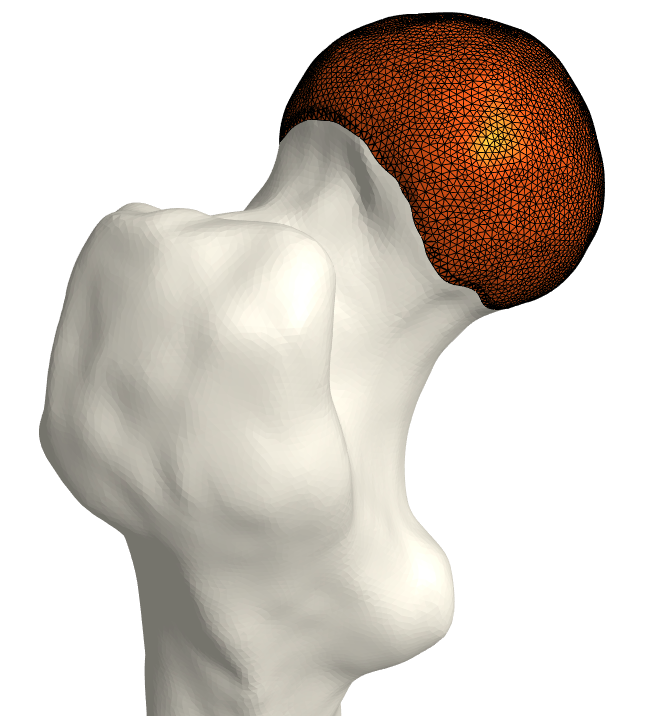}
     \caption{Femoral cartilage.}
     \end{subfigure}
    \hspace{1cm}
    \begin{subfigure}[b]{0.25\linewidth}
        \centering
        \includegraphics[width=\linewidth]{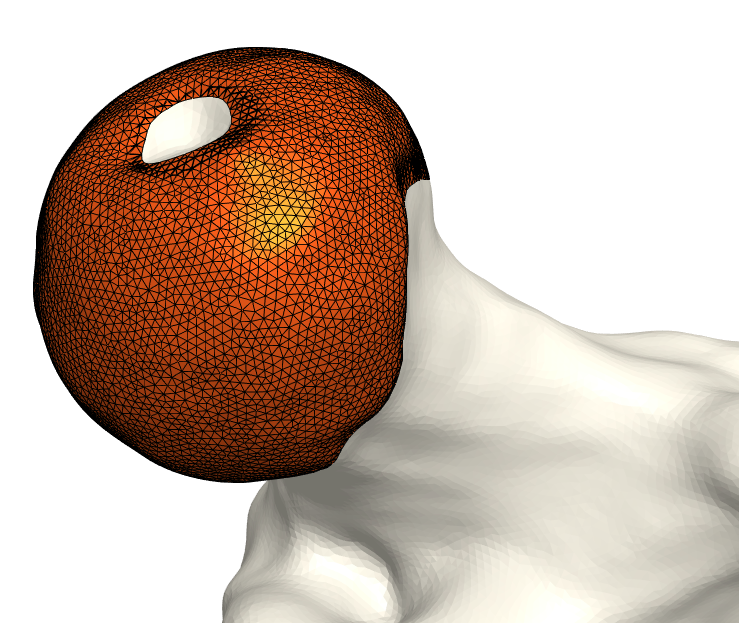}
        \caption{Femoral cartilage.}
    \end{subfigure}
    \hspace{1cm}
    \begin{subfigure}[b]{0.25\linewidth}
        \centering
        \includegraphics[width=.8\linewidth]{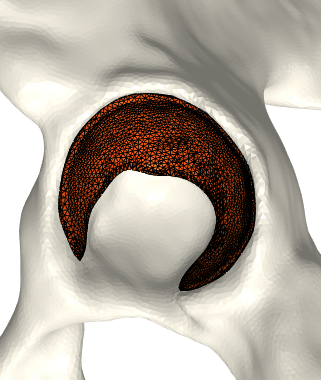}
        \caption{Pelvic cartilage.}
    \end{subfigure}
    \caption{The final cartilage generated by our method for a femur \textbf{(a), (b)} and a pelvis \textbf{(c)}. Notice how the cartilage aligns with the anatomical lines.}
    \label{fig:final}
\end{figure}

The final step of the cartilage generation is to connect the bone-attached region, $(\vertices^{D}_{C}, \faces^{D}_{C})$, and the extruded surface, $(\vertices_{C}^{E}, \faces^{E}_{C})$.
Referring back to Section~\ref{sec:extrusion} we know that $\faces_{C}^{E} \subset \faces^{D}_{C}$.
We create a copy of the set $\faces_{C}^{D} \setminus \faces_{C}^{E}$ and denote it as $\faces^{H}_{C}$.
This new set will connect the bone-attached region to the extruded surface and consists of the faces from the bone-attached region not initially selected for extrusion.
To create a smooth blend between the disjoint surfaces, we apply a \textit{biharmonic weighting} scheme.
We compute the blended extrusion thickness between the two regions by minimizing the Laplacian energy on the boundary of the domain, $\Gamma$.
The extrusion thickness is then the minimizers of,
\begin{align}
    \arg\min\limits_{\vec w} \sum\limits_{w_j \in \vec w} \frac{1}{2}\int\limits_{\Gamma} \lVert \Delta w_j \rVert^2 dV
\end{align}
subject to constraints enforcing interpolation between the boundary of the bone-attached region and the boundary of the extruded region.
The Laplacian energy is discretized using the FEM method and subsequently solved using an active set method~\cite{nocedal2006numerical}.
For more details, we refer the reader to~\cite{jacobson2011bounded}.
The resulting displacements $\vec w$ are then applied to the vertices of $\vertices^{H}_{C}$ as,
\begin{align}
    \vec v_{j} \leftarrow \vec v_{j} + \vec n_{j} w_j, \quad \forall \vec v_{j} \in \vertices^{H}_{C}
    \,.
\end{align}
That is, we extrude each vertex $\vec v_j$ in the direction of its normal by the displacement $w_j$. As a final step, we invert the bone-attached face normals before collecting the three disjoint sets of faces and vertices into a single mesh, $(\vertices_{C}, \faces_{C})$. Observe in Fig.~\ref{fig:final} the cartilage sub-surfaces combined into a single mesh.

\section{Results} \label{sec:results}
We use five CT scans from the TCGA-BLCA collection publicly available at the Cancer Imaging Archive \cite{clark2013cancer}. These subjects are in a supine position and have no diagnosed disease related to the HJ. Each dataset contains a left and right HJ summing up to ten HJs. We generate pelvic and femoral cartilages for all ten joints. 

Our algorithm is implemented in Python using the Libigl library~\cite{libigl}. To quantify the speed of our method, we timed the generation of femoral and pelvic cartilages for all ten joints. All cartilages are generated on a MacBook Pro 2018 with a 2.7 GHz quad-core Intel i7. The initial bone geometries for the ten HJs consist of a similar number of triangles (65k triangles on average). We include all steps described in Section~\ref{sec:method} in the computation time and exclude the time needed for segmentation and preprocessing of the bone models.
We observe that, on average, femoral cartilage reconstruction takes 39.90s and pelvic cartilage reconstruction takes 12.82s. These results show that the time needed to construct the femoral cartilage is significantly higher than for the pelvic cartilages - a factor of three times higher. The femoral cartilage is the only cartilage where the initially trimmed layer (shown in Figure~\ref{fig:initial:guess:femur:trimming}) needs to be grown, explaining the difference in time.
It should be noted that while generating a single piece of cartilage takes less than a minute, generating a \textit{good} model of the cartilage takes significantly longer.
The time needed to generate such cartilage model includes the time required to calibrate the free parameters of our model.
In our experience, the parameters for one patient generally translate well to other models, which means that only minimal tuning is required once a good set of parameter values have been found for a single patient. See supplementary material for the parameters used to generate the cartilage models.

\begin{figure}[ht]
    \centering
    \begin{subfigure}[b]{0.4\linewidth}
        \centering
        \includegraphics[height=0.75\linewidth]{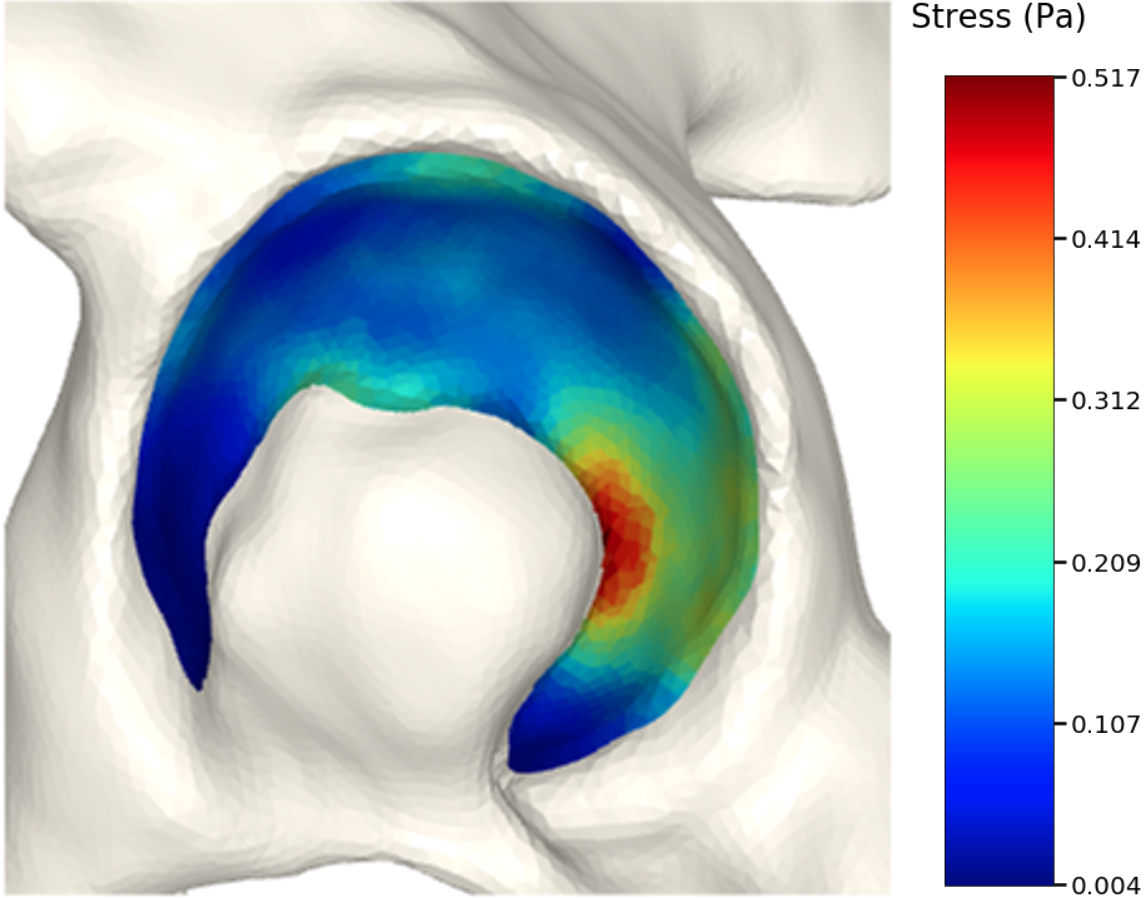}
        \caption{Von Mises stress.}
    \end{subfigure}
    \hspace{1cm}
    \begin{subfigure}[b]{0.4\linewidth}
        \centering
        \includegraphics[height=0.75\linewidth]{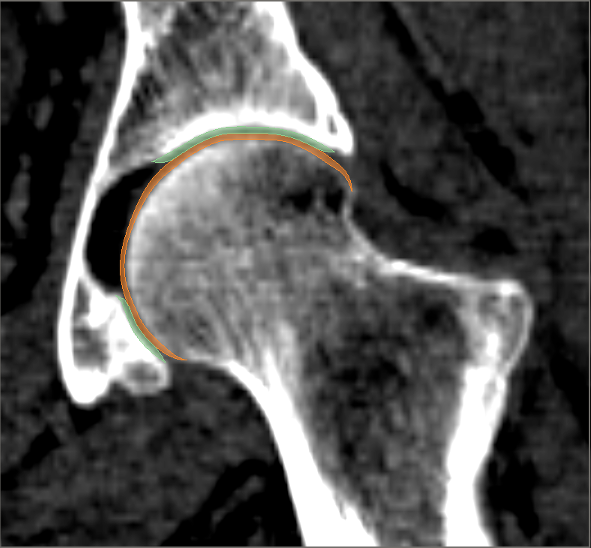}
        \caption{CT scan overlay.}
    \end{subfigure}
    \caption{
    The von Mises stress patterns (\textbf{(a)}) and the generated cartilage imposed on the CT scan from which the bone was extracted (\textbf{(b)}).
    Notice the high level of congruence in the cartilage-bone interfaces and cartilage-cartilage interface.}
    \label{fig:von:mises:stress}
\end{figure}

We have qualitatively verified that the articulating surfaces in all the ten HJs are detected correctly regardless of their anatomical variance using visual inspection of overlays as shown in Fig.~\ref{fig:von:mises:stress}. As desired, we observe a high degree of congruence between the opposing joint surfaces, meaning no gaps or overlaps in the cartilage-cartilage interface. Moreover, we observe a smooth transition towards the bone geometries as expected from the correct anatomy. The parameter values used to generate the cartilage from Fig.~\ref{fig:future:joints:hip}. The free parameters are the neighbourhood-size used to estimate the curvature of the bone ($\mathcal{N}$); the minimum and maximum curvature in the cartilage region ($\kappa_{\min}, \kappa_{\max}$; Eq.~\eqref{eq:grow}); the distance parameter in $mm$ ($\delta$; Eq.~\ref{eq:distance:filtering}); and the number of times the outer boundary should be trimmed ($N_{trim}$). Here, the curvature based parameters ($\mathcal{N}, \kappa_{\min}, \kappa_{\max})$ are only used for the femur. See supplementary material for more visual comparisons.

Further, we quantitatively compare values of our cartilage's geometric measures to data obtained from the literature to assess that we agree with state-of-the-art reported values. Results are in Table~\ref{table:comparisons}. We observe that our method agrees with the range of values from both manual and semi-automated approaches~\cite{anderson2008validation,c15,lazik2016,schmaranzer2019automatic}.

\begin{table}
\centering
\caption{Geometric measure comparison between our method and literature}
\begin{tabular}{ l | c | c | c | c}
    \multicolumn{5}{l}{} \\
    Measure &
    Study &
    \# of hip joints &
    Femur &
    Pelvis  
    \\
\hline
    \multirow{5}{0.2\linewidth}{Mean thickness (mm)} 
    & Our                             & 10 & $0.81 \pm 0.08 $ & $ 1.05 \pm 0.06 $ \\
    & \cite{anderson2008validation}   & 1  & $1.5\pm0.5$            & $1.6\pm0.4$             \\
    & \cite{c15}                      & 26 & $1.18 \pm 0.06 $ & $1.26 \pm 0.04$   \\
    & \cite{lazik2016}                & 11 &          1.0       & 1.8                \\
    & \cite{schmaranzer2019automatic} & 23 &            -      & $3.5 \pm 0.9$      \\
    & \cite{c3}                       & -  & 1.28 & 1.28                \\
\hline
    \multirow{2}{0.2\linewidth}{Bone coverage area($mm^2$)}
    & Our                             & 10 & $5349.41 \pm 660.51$ & $2848.98 \pm 351.86$ \\
    & \cite{schmaranzer2019automatic} & 23 &           -           & $1634 \pm 400$       \\
 \hline
  \multirow{2}{0.2\linewidth}{Bone coverage percentage }
    & Our        & 10  &  - & $39\% \pm 7$  \\
    & \cite{c14} & 16   &  - & 34\%          \\
   \hline
  Contact area ($mm^2$) & Our & 10 &&$1732.01 \pm 281.06 $\\
\end{tabular}
 \label{table:comparisons}
\end{table}

Next, we analyze the simulation quality of the generated models in an FE analysis setup. We use the Tetgen package to generate a volumetric mesh with tetrahedral elements for each geometry\cite{tetgen}. Using a displacement-controlled simulation in the FEBio software package \cite{maas2012febio}, we push the pelvis on top of the femur, representing a pseudo stance position. Considering the stance position, the distal femur is fixed in the x,y, and z-directions. The pelvis is moved in the z-direction by $1mm$ towards the femur. We simplified the mechanical behavior of all the tissues to isotropic and linearly elastic. The material properties are based on the review in \cite{R13}.
The bone-cartilage interfaces are modeled as \textit{Tied contacts}. An augmented surface contact algorithm with friction-less tangential interaction is applied to the cartilage-cartilage interfaces allowing unhindered motion in the HJ.

Fig.~\ref{fig:von:mises:stress} visualizes the von Mises stress pattern on the pelvic cartilage for one HJ. More are shown in the supplementary material. We have verified that no spurious stress peaks appear and that stress values change gradually and smoothly across the cartilage. Further, the high-stress areas are located in the up-direction, as we expect from the applied displacements. The stress values and patterns are not to be confused with those from a real stance. They only serve as a verification test of simulation properties. For such a simulation, we require ligaments and muscles to stabilize the girdle and a correction from sublime pose bias.

\section{Discussion and Conclusion} \label{sec:discussion}

The results show our method produces similar cartilage to manual segmentation and adheres to our clinical assumptions about cartilage morphology~\cite{standring2020gray}. The high congruence level prevents potential spikes and peak stresses in the articulating surfaces, resulting in continuous stress distribution in the HJ. 

Our subjects are in a supine position, which means the cartilage is under a horizontal load. In this case, the bones are not in an unloaded position, causing the generated cartilages to be in a pre-load state. The supine position explains the difference between the cartilage thickness and contact area compared to other studies. One solution is to relocate the bone geometries to the desired position before generating the cartilages or optimizing the rest shape used in the FE analysis. Moreover, as we measure the contact area before running any simulations, our results differ from other studies reporting this value as a simulation result in different gait cycles.

As we rely on the bone curvatures for cartilage-bone interface definition, accurate bone geometries are crucial for realistic results. If the segmented bones are too smooth or coarse, the final product will not agree with real cartilage anatomy. We want to investigate further detailed geometric validation of our method; however, no public databases contain segmented cartilage for hip joints to the best of our knowledge. We leave this for future work.

\begin{figure}
    \centering
    \begin{subfigure}[b]{0.6\linewidth}
        \centering
        \includegraphics[width=\linewidth]{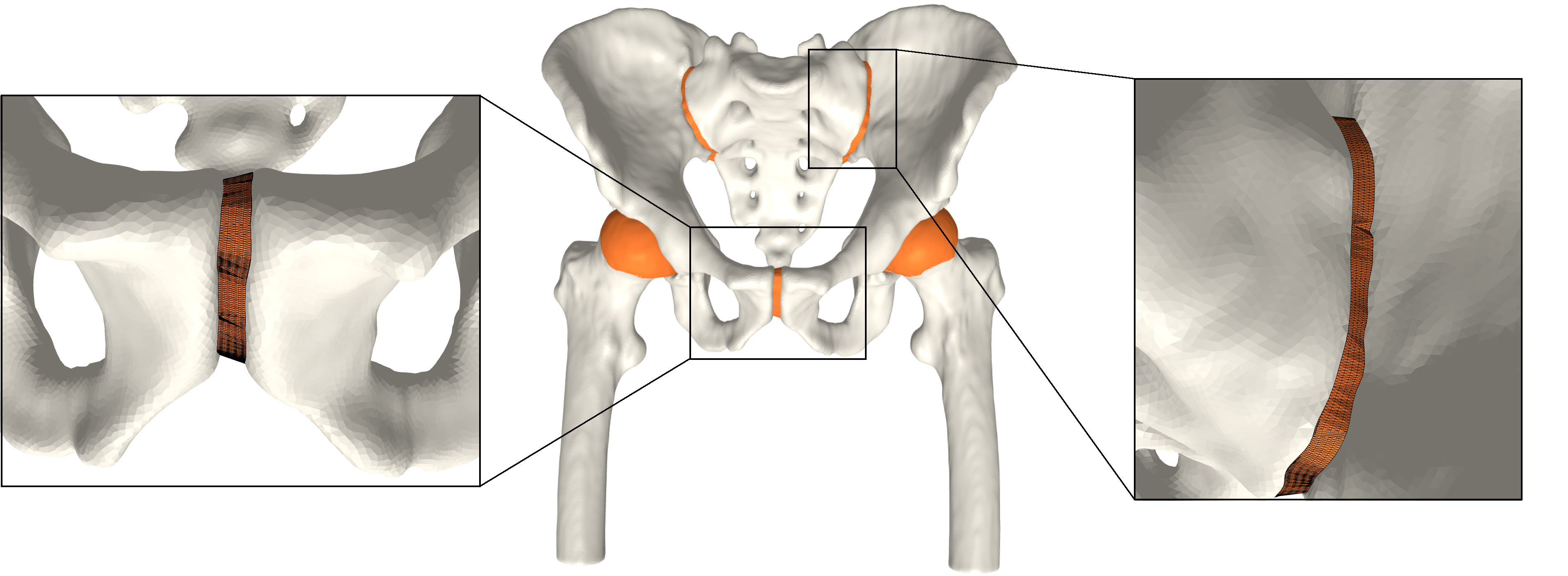}
        \caption{}
        \label{fig:future:joints:hip}
    \end{subfigure}
    \begin{subfigure}[b]{0.25\linewidth}
        \centering
        \includegraphics[width=\linewidth]{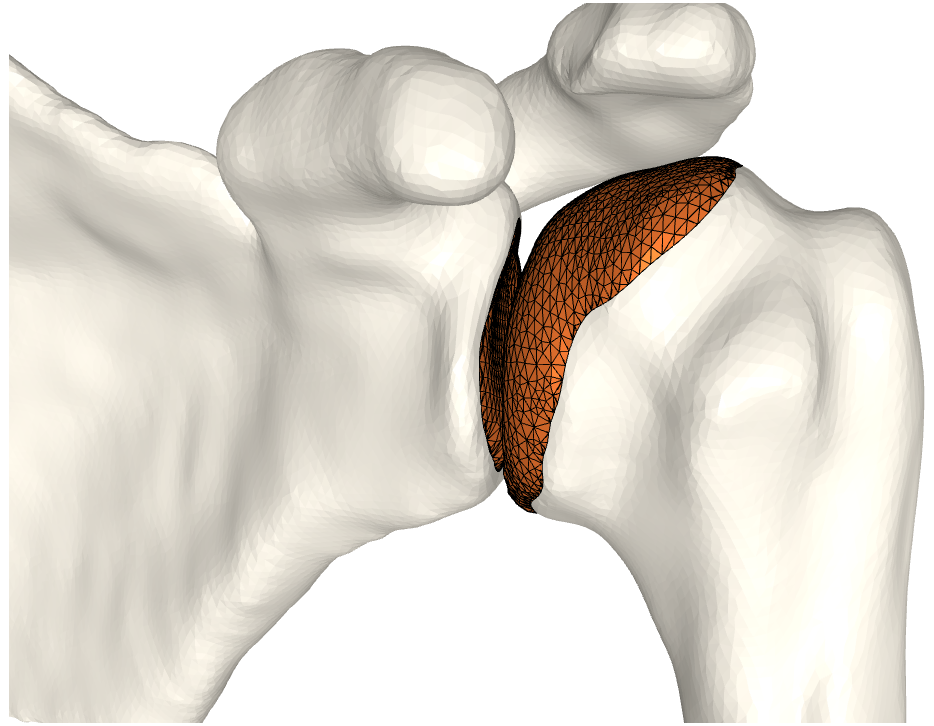}
        \caption{}
    \end{subfigure}
    \begin{subfigure}[b]{0.125\linewidth}
        \centering
        \includegraphics[width=\linewidth]{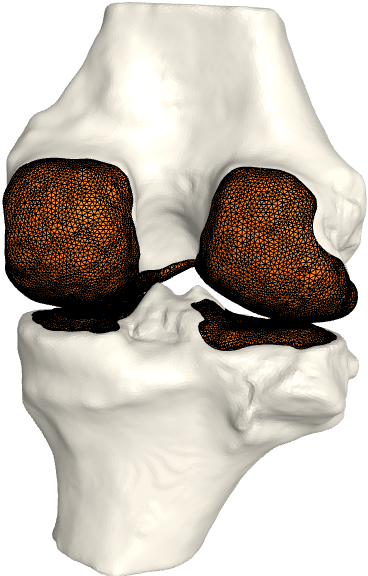}        \caption{}
    \end{subfigure}
    \caption{\textbf{a)} All generated cartilages. Zoom boxes highlight the pubic and sacroiliac joint cartilages.
    \textbf{b)} Shoulder joint cartilage \textbf{c)} Femoral and tibial cartilage in the knee joint. Bone models are acquired from available datasets \cite{clark2013cancer,dzialo2018development,dzialo2019evaluation}.}
    \label{fig:future:joints}
\end{figure}

The FE analysis results show that the generated models produce smooth stress patterns in a pseudo stance without any geometry-related convergence issues. As mentioned in section~\ref{sec:results}, these results only serve as a verification test of simulation properties. We need a more advanced simulation setting to model a real stance position. We leave this for future work. Moreover, we believe that minor modifications will estimate the shoulder joint cartilage since it has a similar ball-in-socket structure. Fig. \ref{fig:future:joints} shows early evidence of generalization. 
In contrast, other joints, such as the knee joint, are challenging. We leave other joint types for future work.

In conclusion, we have created a method to automate the reconstruction of healthy subject-specific HJ cartilages independent from clinical images. The proposed method is fast and preserves the anatomical properties of the cartilage tissue. The proposed method generates high-quality cartilage without the need for any manual segmentation or training data. We have shown the cartilage models can be used in FE analysis and believe our method will enable large-scale population studies. Future work will extend the geometric principles to abnormal HJs and other joints such as the shoulder and knee joint. Our code is open-source and available from \url{https://github.com/diku-dk/CarGen}.
\section{Acknowledgments}

\begin{wrapfigure}{l}{0.15\linewidth}
\centering
\vspace{-25pt}
 \vspace{-30pt}
\end{wrapfigure}
This project has received funding from the European Union’s Horizon 2020 research and innovation programme under the Marie Sklodowska-Curie grant agreement No. 764644.
This paper only contains the author's views, and the Research Executive Agency and the Commission are not responsible for any use that may be made of the information it contains.\\
%
This project has also received funding from Independent Research Fund Denmark (DFF) under agreement No. 9131-00085B.

\bibliographystyle{splncs04}
\bibliography{references}

\begin{thebibliography}{10}
\providecommand{\url}[1]{\texttt{#1}}
\providecommand{\urlprefix}{URL }
\providecommand{\doi}[1]{https://doi.org/#1}

\bibitem{anderson2008validation}
Anderson, A.E., Ellis, B.J., Maas, S.A., Peters, C.L., Weiss, J.A.: Validation
  of finite element predictions of cartilage contact pressure in the human hip
  joint. Journal of biomechanical engineering  \textbf{130}(5) (2008)

\bibitem{c3}
Anderson, A.E., Ellis, B.J., Maas, S.A., Weiss, J.A.: Effects of idealized
  joint geometry on finite element predictions of cartilage contact stresses in
  the hip. Journal of biomechanics  \textbf{43}(7),  1351--1357 (2010)

\bibitem{clark2013cancer}
Clark, K., Vendt, B., Smith, K., Freymann, J., Kirby, J., Koppel, P., Moore,
  S., Phillips, S., Maffitt, D., Pringle, M., et~al.: The cancer imaging
  archive (tcia): maintaining and operating a public information repository.
  Journal of digital imaging  \textbf{26}(6),  1045--1057 (2013)

\bibitem{dzialo2019evaluation}
Dzialo, C.M., Pedersen, P.H., Jensen, K.K., de~Zee, M., Andersen, M.S.:
  Evaluation of predicted patellofemoral joint kinematics with a moving-axis
  joint model. Medical engineering \& physics  \textbf{73},  85--91 (2019)

\bibitem{dzialo2018development}
Dzialo, C., Pedersen, P., Simonsen, C., Jensen, K., de~Zee, M., Andersen, M.:
  Development and validation of a subject-specific moving-axis tibiofemoral
  joint model using mri and eos imaging during a quasi-static lunge. Journal of
  biomechanics  \textbf{72},  71--80 (2018)

\bibitem{c12}
Gilles, B., Magnenat-Thalmann, N.: Musculoskeletal mri segmentation using
  multi-resolution simplex meshes with medial representations. Medical image
  analysis  \textbf{14}(3),  291--302 (2010)

\bibitem{c14}
Harris, M.D., Anderson, A.E., Henak, C.R., Ellis, B.J., Peters, C.L., Weiss,
  J.A.: Finite element prediction of cartilage contact stresses in normal human
  hips. Journal of Orthopaedic Research  \textbf{30}(7),  1133--1139 (2012)

\bibitem{c16}
Henak, C.R., Anderson, A.E., Weiss, J.A.: Subject-specific analysis of joint
  contact mechanics: application to the study of osteoarthritis and surgical
  planning. Journal of biomechanical engineering  \textbf{135}(2) (2013)

\bibitem{c2}
Henak, C.R., Carruth, E.D., Anderson, A.E., Harris, M.D., Ellis, B.J., Peters,
  C.L., Weiss, J.A.: Finite element predictions of cartilage contact mechanics
  in hips with retroverted acetabula. Osteoarthritis and cartilage
  \textbf{21}(10),  1522--1529 (2013)

\bibitem{c8-2}
Ipavec, M., Brand, R., Pedersen, D., Mav{\v{c}}i{\v{c}}, B., Kralj-Igli{\v{c}},
  V., Igli{\v{c}}, A.: Mathematical modelling of stress in the hip during gait.
  Journal of Biomechanics  \textbf{32}(11),  1229--1235 (1999)

\bibitem{c8-1}
Ipavec, M., Igli{\v{c}}, A., Igli{\v{c}}, V.K., Srakar, F.: Stress distribution
  on the hip joint articular surface during gait. Pfl{\"u}gers Archiv
  \textbf{431}(6),  R275--R276 (1996)

\bibitem{jacobson2011bounded}
Jacobson, A., Baran, I., Popovic, J., Sorkine, O.: Bounded biharmonic weights
  for real-time deformation. ACM Trans. Graph.  \textbf{30}(4), ~78 (2011)

\bibitem{libigl}
Jacobson, A., Panozzo, D., et~al.: {libigl}: A simple {C++} geometry processing
  library (2018), https://libigl.github.io/

\bibitem{jorge2014finite}
Jorge, J., Sim{\~o}es, F., Pires, E., Rego, P., Tavares, D., Lopes, D., Gaspar,
  A.: Finite element simulations of a hip joint with femoroacetabular
  impingement. Computer methods in biomechanics and biomedical engineering
  \textbf{17}(11),  1275--1284 (2014)

\bibitem{lazik2016}
Lazik, A., Theysohn, J.M., Geis, C., Johst, S., Ladd, M.E., Quick, H.H., Kraff,
  O.: 7 tesla quantitative hip mri: T1, t2 and t2* mapping of hip cartilage in
  healthy volunteers. European radiology  \textbf{26}(5),  1245--1253 (2016)

\bibitem{c1}
Li, F., Chen, H., Mawatari, T., Iwamoto, Y., Jiang, F., Chen, X.: Influence of
  modeling methods for cartilage layer on simulation of periacetabular
  osteotomy using finite element contact analysis. Journal of Mechanics in
  Medicine and Biology  \textbf{18}(02),  1850018 (2018)

\bibitem{c9}
Lindner, C., Thiagarajah, S., Wilkinson, J.M., Wallis, G.A., Cootes, T.F.,
  arcOGEN Consortium, et~al.: Accurate bone segmentation in 2d radiographs
  using fully automatic shape model matching based on regression-voting. In:
  International Conference on Medical Image Computing and Computer-Assisted
  Intervention. pp. 181--189. Springer (2013)

\bibitem{lopes2020hip}
Lopes, D.S., Pires, S.M., Barata, C.D., Mascarenhas, V.V., Jorge, J.A.: The hip
  joint as an egg shape: a comprehensive study of femoral and acetabular
  morphologies. Computer Methods in Biomechanics and Biomedical Engineering:
  Imaging \& Visualization  \textbf{8}(4),  411--425 (2020)

\bibitem{maas2012febio}
Maas, S.A., Ellis, B.J., Ateshian, G.A., Weiss, J.A.: Febio: finite elements
  for biomechanics. Journal of biomechanical engineering  \textbf{134}(1),
  011005 (2012)

\bibitem{c15}
Mechlenburg, I., Nyengaard, J.R., Gelineck, J., Soballe, K.: Cartilage
  thickness in the hip joint measured by mri and stereology--a methodological
  study. Osteoarthritis and cartilage  \textbf{15}(4),  366--371 (2007)

\bibitem{c5}
Menschik, F.: The hip joint as a conchoid shape. Journal of Biomechanics
  \textbf{30}(9),  971--973 (1997)

\bibitem{nocedal2006numerical}
Nocedal, J., Wright, S.: Numerical optimization. Springer Science \& Business
  Media (2006)

\bibitem{panozzo2010efficient}
Panozzo, D., Puppo, E., Rocca, L.: Efficient multi-scale curvature and crease
  estimation. Proceedings of Computer Graphics, Computer Vision and Mathematics
  (Brno, Czech Rapubic  \textbf{1}(6) (2010)

\bibitem{c7}
Russell, M.E., Shivanna, K.H., Grosland, N.M., Pedersen, D.R.: Cartilage
  contact pressure elevations in dysplastic hips: a chronic overload model.
  Journal of orthopaedic surgery and research  \textbf{1}(1), ~6 (2006)

\bibitem{schmaranzer2019automatic}
Schmaranzer, F., Helfenstein, R., Zeng, G., Lerch, T.D., Novais, E.N., Wylie,
  J.D., Kim, Y.J., Siebenrock, K.A., Tannast, M., Zheng, G.: Automatic
  mri-based three-dimensional models of hip cartilage provide improved
  morphologic and biochemical analysis. Clinical orthopaedics and related
  research  \textbf{477}(5), ~1036 (2019)

\bibitem{tetgen}
Si, H.: Tetgen, a delaunay-based quality tetrahedral mesh generator. ACM Trans.
  Math. Softw.  \textbf{41}(2),  11:1--11:36 (feb 2015)

\bibitem{standring2020gray}
Standring, S.: Gray's anatomy e-book: the anatomical basis of clinical
  practice. Elsevier (2020)

\bibitem{c6}
Tsumura, H., Kaku, N., Ikeda, S., Torisu, T.: A computer simulation of
  rotational acetabular osteotomy for dysplastic hip joint: does the optimal
  transposition of the acetabular fragment exist? Journal of Orthopaedic
  Science  \textbf{10}(2),  145--151 (2005)

\bibitem{R13}
Vafaeian, B., Zonoobi, D., Mabee, M., Hareendranathan, A., El-Rich, M., Adeeb,
  S., Jaremko, J.: Finite element analysis of mechanical behavior of human
  dysplastic hip joints: a systematic review. Osteoarthritis and cartilage
  \textbf{25}(4),  438--447 (2017)

\bibitem{c11}
Van~Houcke, J., Audenaert, E.A., Atkins, P.R., Anderson, A.E.: A combined
  geometric morphometric and discrete element modeling approach for hip
  cartilage contact mechanics. Frontiers in Bioengineering and Biotechnology
  \textbf{8} (2020)

\bibitem{c10}
Yokota, F., Okada, T., Takao, M., Sugano, N., Tada, Y., Tomiyama, N., Sato, Y.:
  Automated ct segmentation of diseased hip using hierarchical and conditional
  statistical shape models. In: International Conference on Medical Image
  Computing and Computer-Assisted Intervention. pp. 190--197. Springer (2013)

\bibitem{c8}
Yoshida, H., Faust, A., Wilckens, J., Kitagawa, M., Fetto, J., Chao, E.Y.S.:
  Three-dimensional dynamic hip contact area and pressure distribution during
  activities of daily living. Journal of biomechanics  \textbf{39}(11),
  1996--2004 (2006)

\end{thebibliography}
\newpage
\setcounter{page}{1}
\section*{Supplementary Material}
\setcounter{figure}{0}

\begin{figure}[h!]
    \centering
    \begin{subfigure}[b]{0.175\linewidth}
        \centering
        \includegraphics[width=\linewidth]{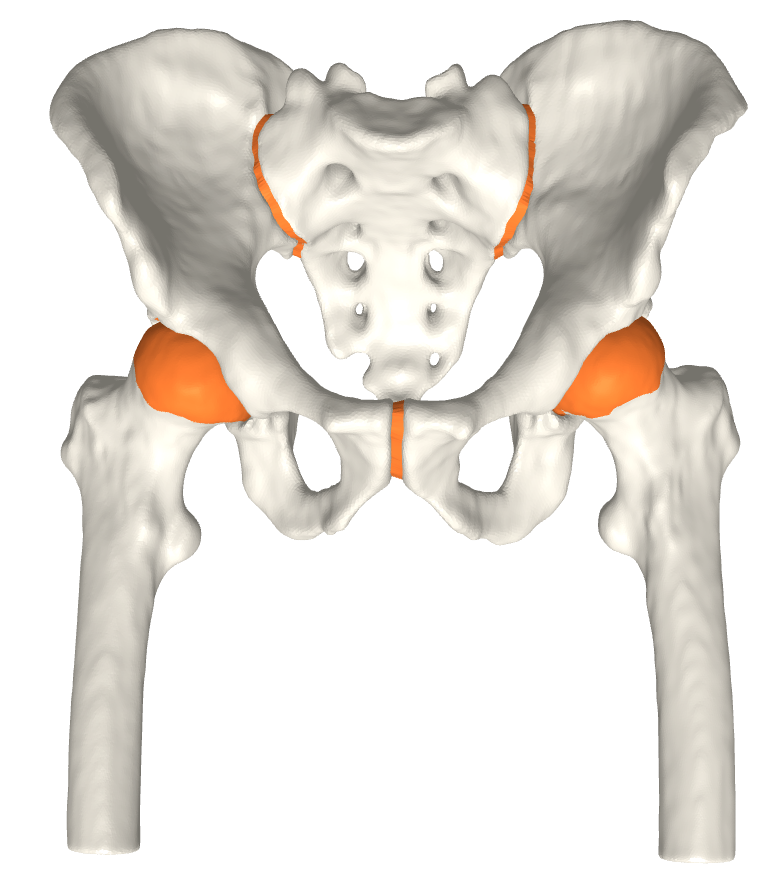}
        \caption{Subject 0}
    \end{subfigure}
    \hspace{1cm}
    \begin{subfigure}[b]{0.175\linewidth}
        \centering
        \includegraphics[width=\linewidth]{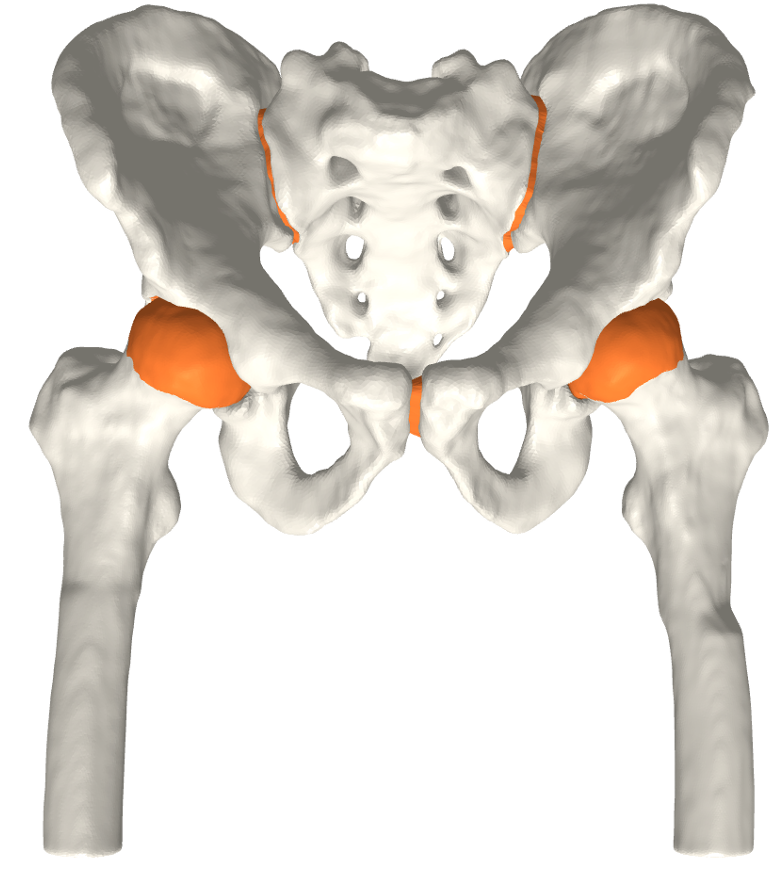}
        \caption{Subject 1}
    \end{subfigure}
    \hspace{1cm}
    \begin{subfigure}[b]{0.175\linewidth}
        \centering
        \includegraphics[width=\linewidth]{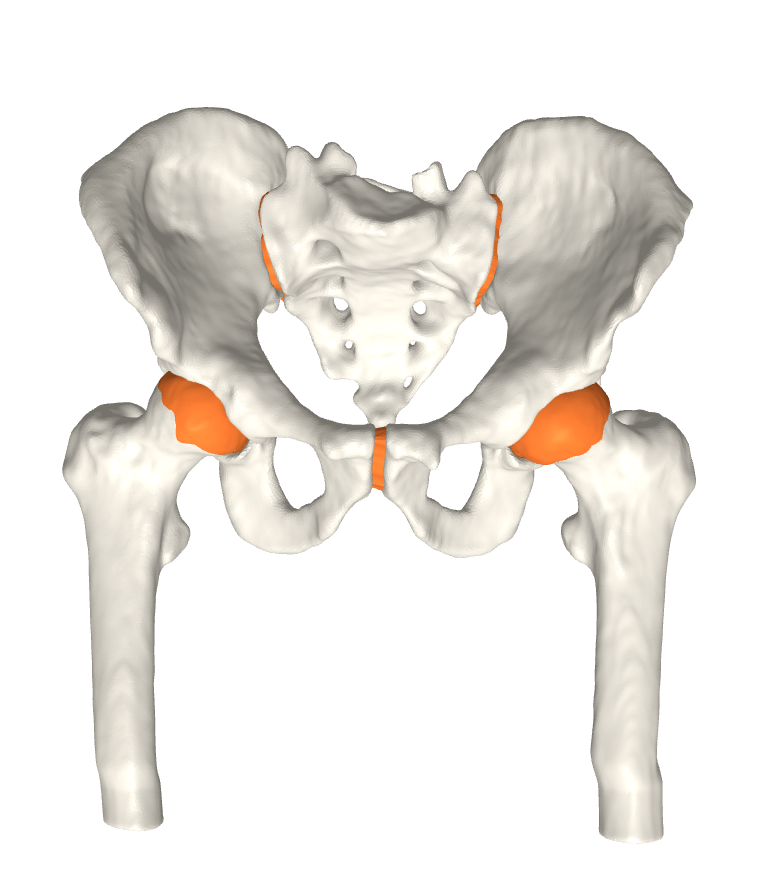}
        \caption{Subject 2}
    \end{subfigure}
    \\
    \begin{subfigure}[b]{0.175\linewidth}
        \centering
        \includegraphics[width=\linewidth]{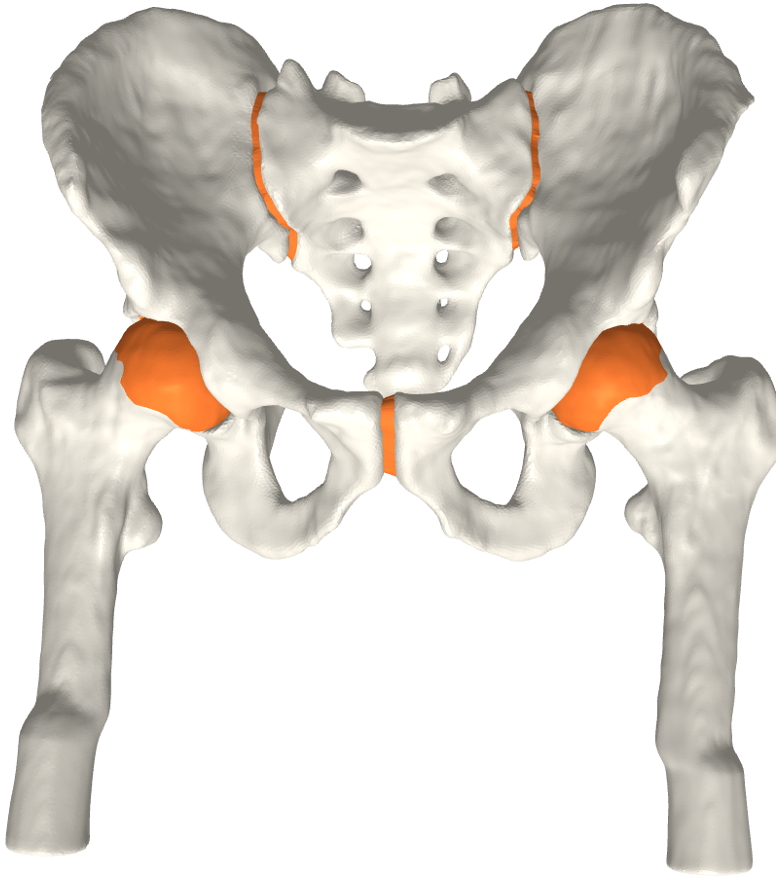}
        \caption{Subject 3}
    \end{subfigure}
    \hspace{1cm}
    \begin{subfigure}[b]{0.175\linewidth}
        \centering
        \includegraphics[width=\linewidth]{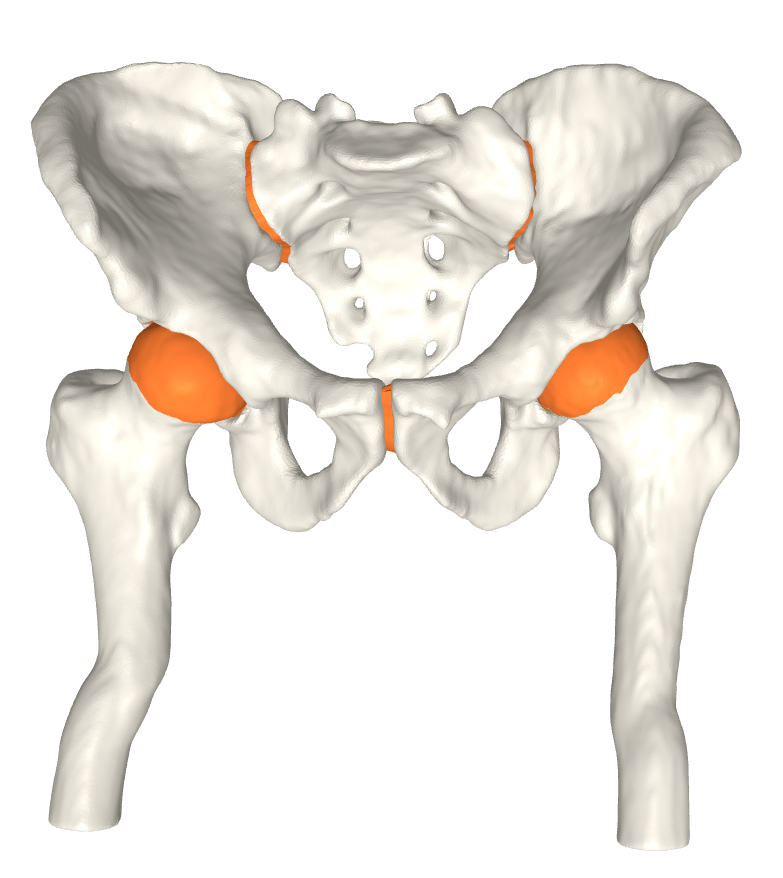}
        \caption{Subject 4}
    \end{subfigure}
    \caption{Cartilage generated for the five subjects in our test set.}
    \label{fig:generated:hip:cartilage:sm}
\end{figure}

\begin{figure}[h!]
    \centering
    \begin{subfigure}[b]{0.175\linewidth}
        \centering
        \includegraphics[width=\linewidth]{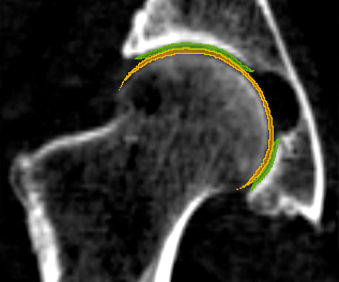}
        \caption{S. 0 right}
    \end{subfigure}
    \begin{subfigure}[b]{0.175\linewidth}
        \centering
        \includegraphics[width=\linewidth]{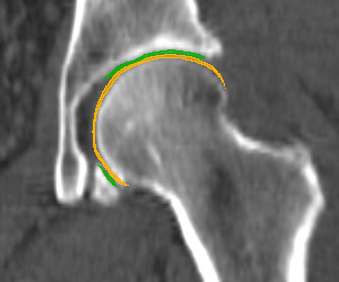}
        \caption{S. 1 left}
    \end{subfigure}
    \begin{subfigure}[b]{0.175\linewidth}
        \centering
        \includegraphics[width=\linewidth]{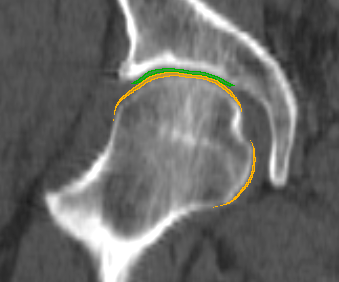}
        \caption{S. 1 right}
    \end{subfigure}
    
    \begin{subfigure}[b]{0.175\linewidth}
        \centering
        \includegraphics[width=\linewidth]{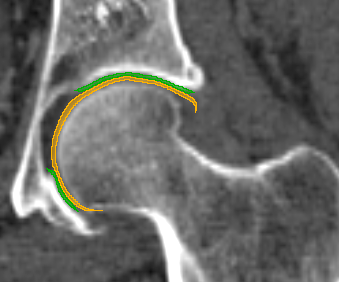}
        \caption{S. 2 left}
    \end{subfigure}
    \begin{subfigure}[b]{0.175\linewidth}
        \centering
        \includegraphics[width=\linewidth]{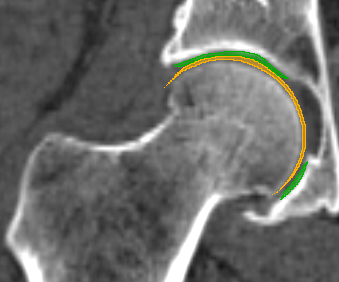}
        \caption{S. 2 right}
    \end{subfigure}
    \begin{subfigure}[b]{0.175\linewidth}
        \centering
        \includegraphics[width=\linewidth]{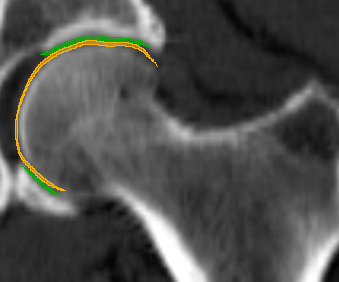}
        \caption{S. 3 left}
    \end{subfigure}
    
    \begin{subfigure}[b]{0.175\linewidth}
        \centering
        \includegraphics[width=\linewidth]{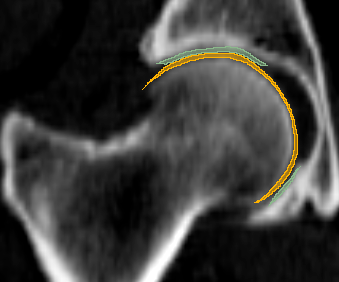}
        \caption{S. 3 right}
    \end{subfigure}
    \begin{subfigure}[b]{0.175\linewidth}
        \centering
        \includegraphics[width=\linewidth]{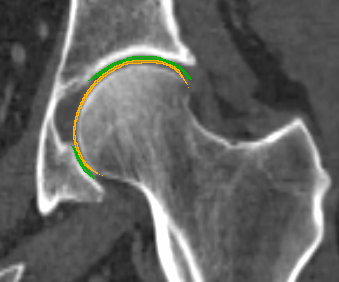}
        \caption{S. 4 left}
    \end{subfigure}
    \begin{subfigure}[b]{0.175\linewidth}
        \centering
        \includegraphics[width=\linewidth]{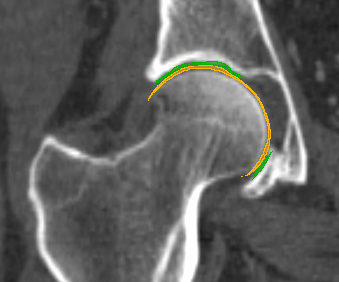}
        \caption{S. 4 right}
    \end{subfigure}
    
    \caption{CT scans overlaid with cartilage and bones for the remaining subjects.}
    \label{fig:overlaid:images:sm}
\end{figure}

\begin{figure}[h!]
    \centering
    \begin{subfigure}[b]{0.3\linewidth}
        \centering
        \includegraphics[width=\linewidth]{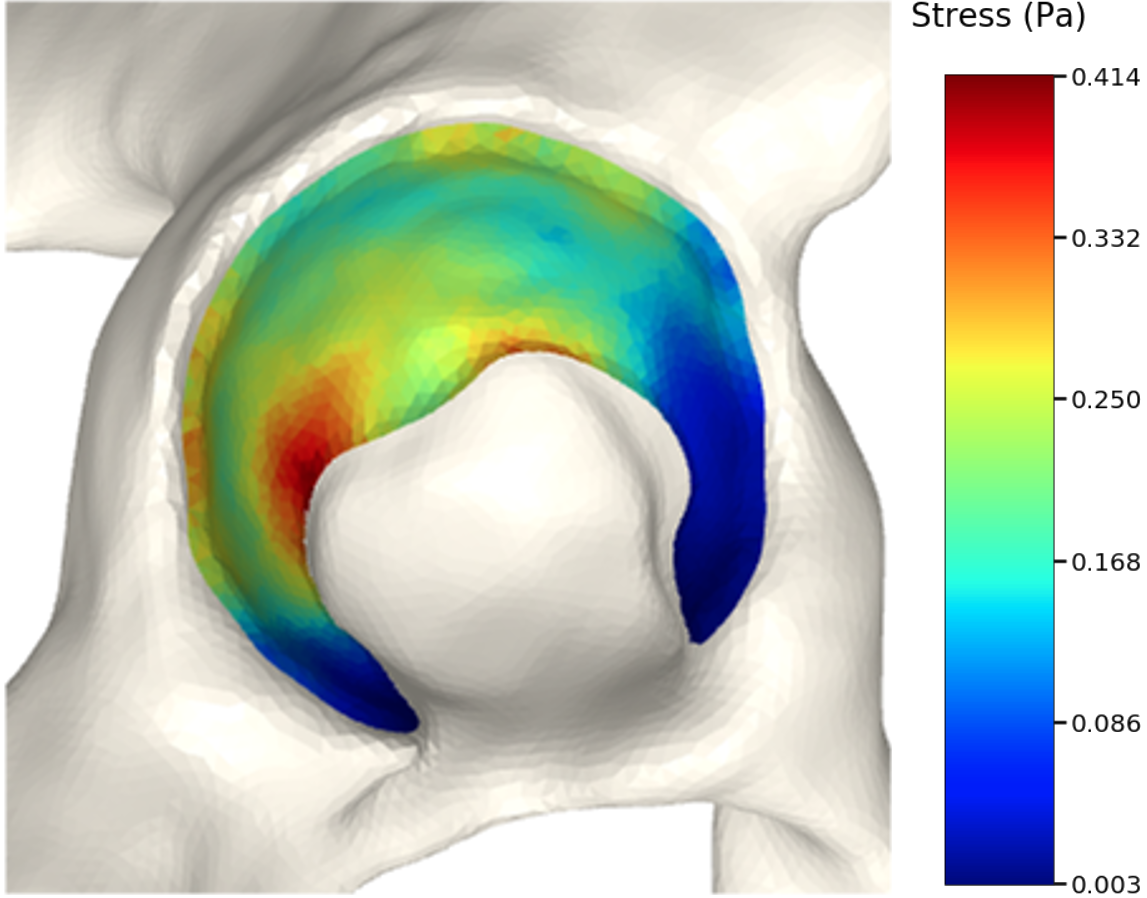}
        \caption{Subject 0 right}
    \end{subfigure}
    \begin{subfigure}[b]{0.3\linewidth}
        \centering
        \includegraphics[width=\linewidth]{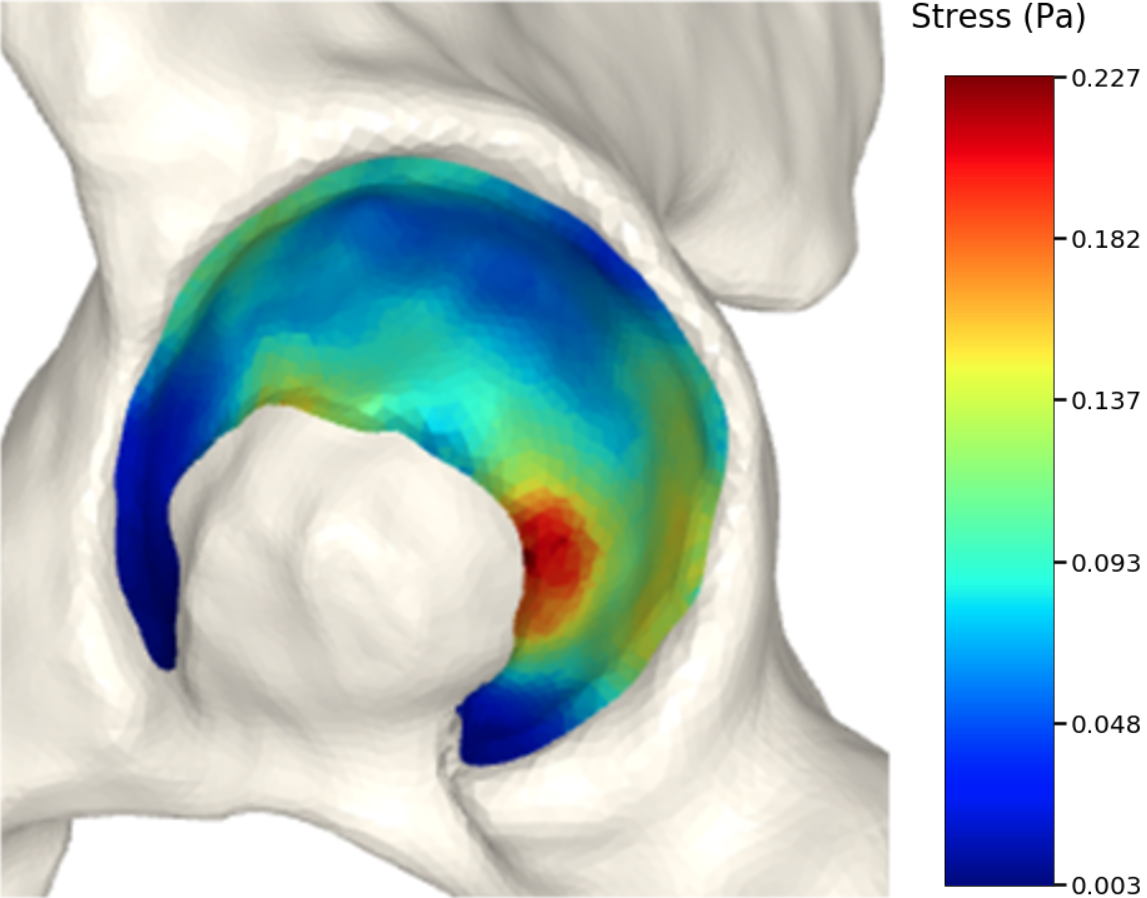}
        \caption{Subject 1 left}
    \end{subfigure}
    \begin{subfigure}[b]{0.3\linewidth}
        \centering
        \includegraphics[width=\linewidth]{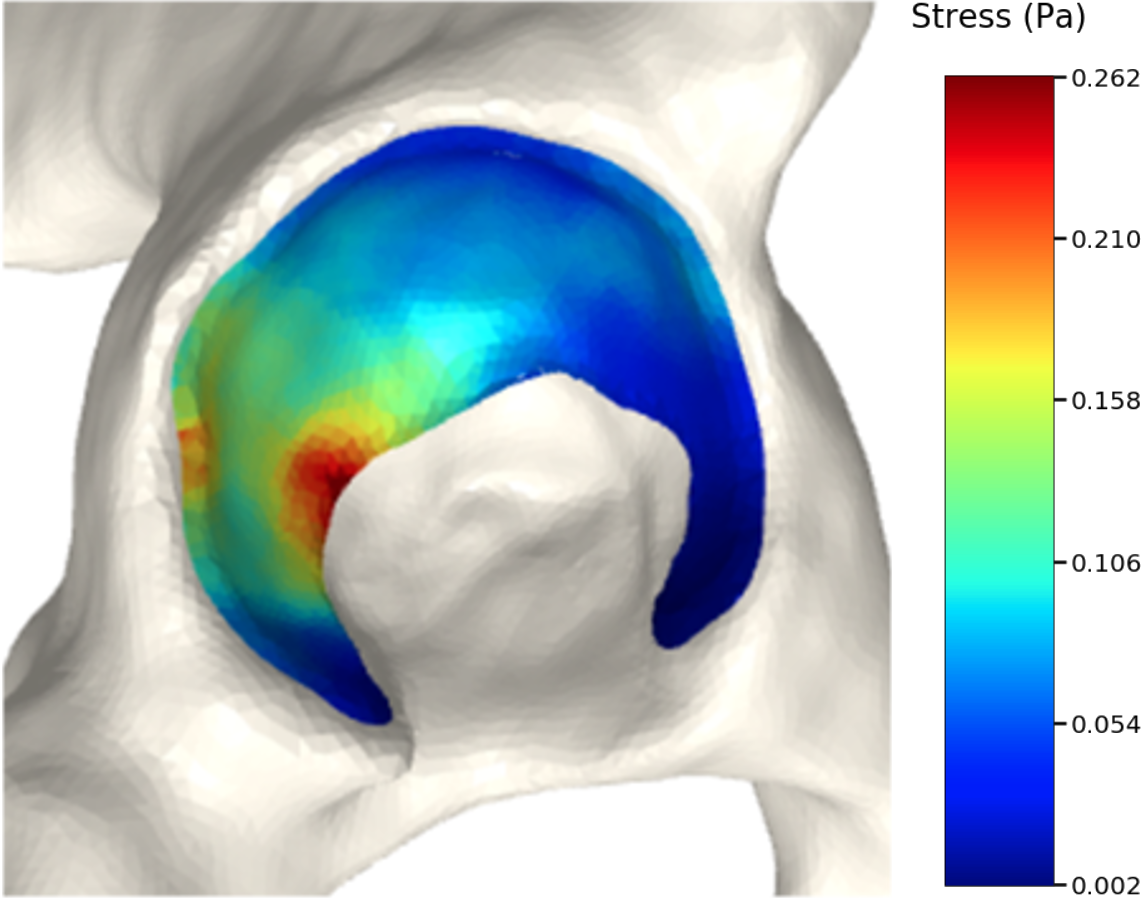}
        \caption{Subject 1 right}
    \end{subfigure}
    
    \begin{subfigure}[b]{0.3\linewidth}
        \centering
        \includegraphics[width=\linewidth]{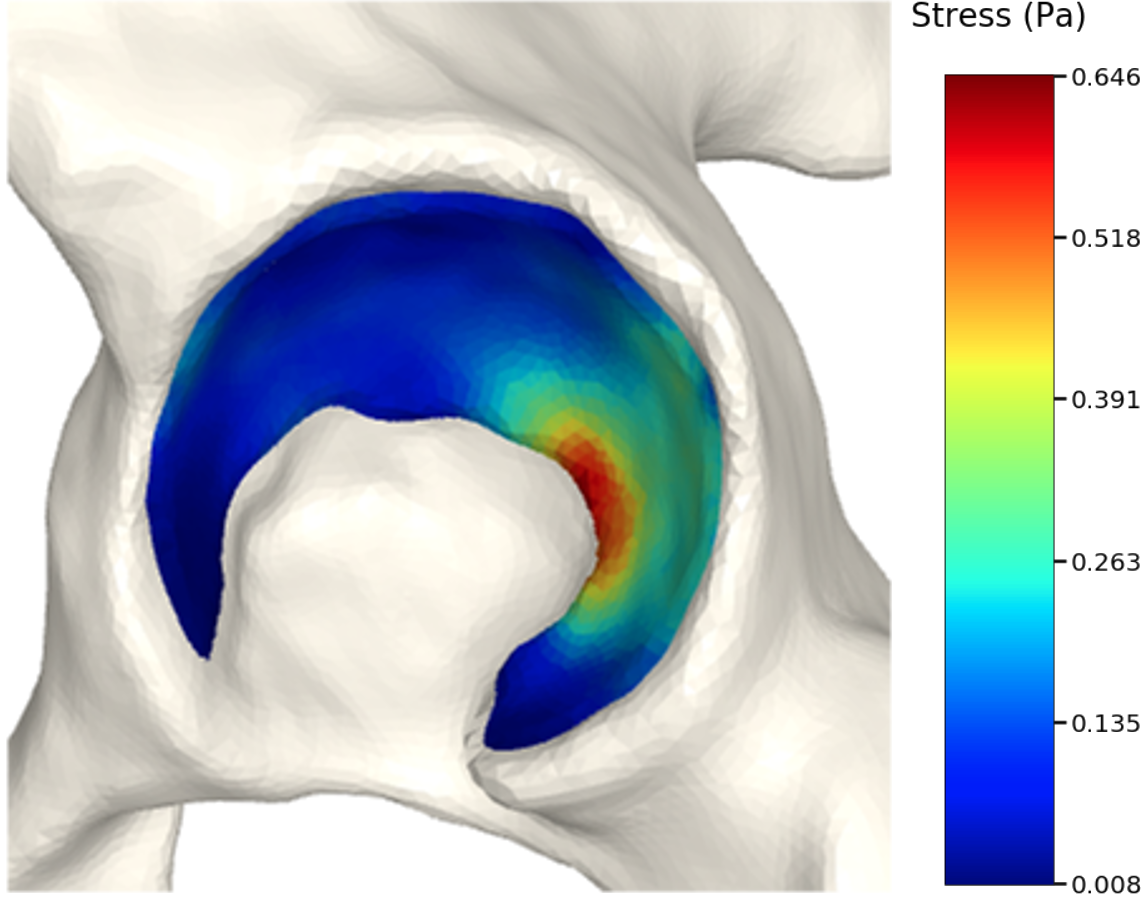}
        \caption{Subject 2 left}
    \end{subfigure}
    \begin{subfigure}[b]{0.3\linewidth}
        \centering
        \includegraphics[width=\linewidth]{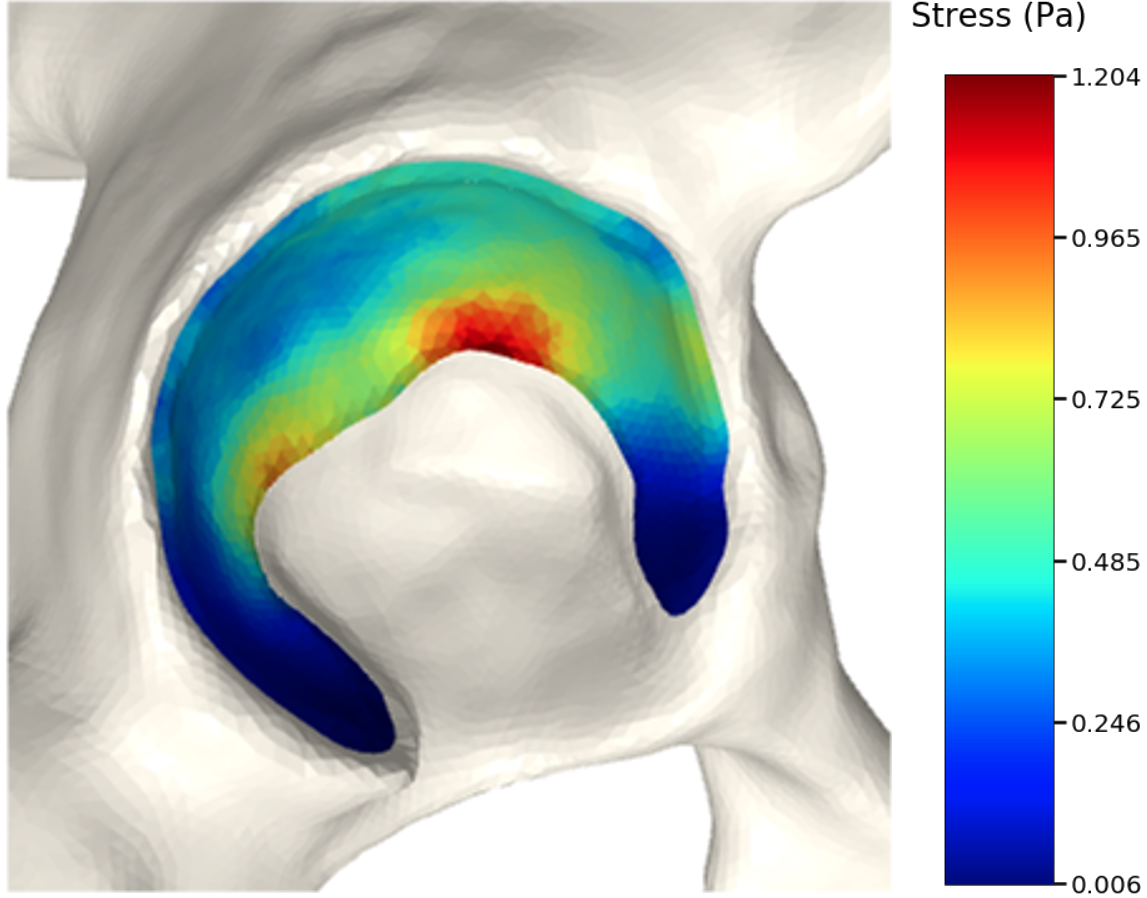}
        \caption{Subject 2 right}
    \end{subfigure}
    \begin{subfigure}[b]{0.3\linewidth}
        \centering
        \includegraphics[width=\linewidth]{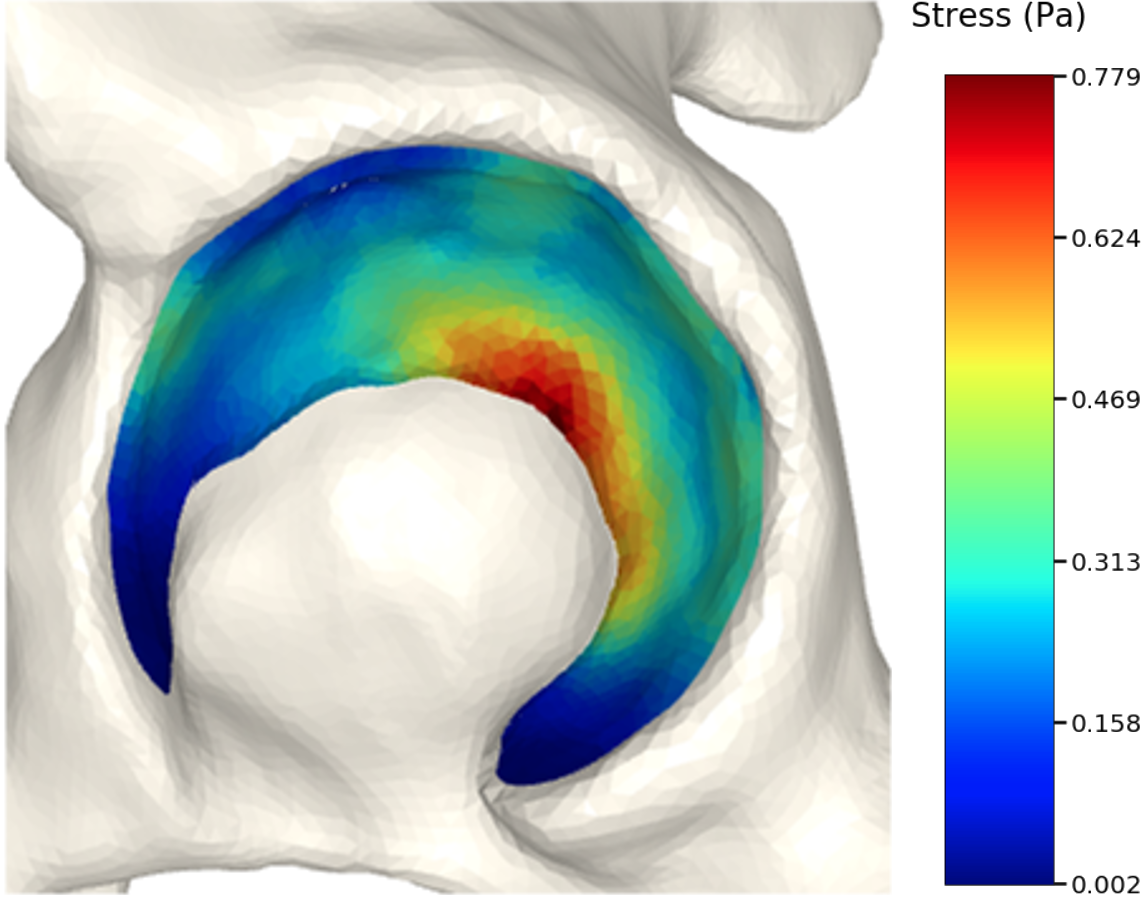}
        \caption{Subject 3 left}
    \end{subfigure}
    
    \begin{subfigure}[b]{0.3\linewidth}
        \centering
        \includegraphics[width=\linewidth]{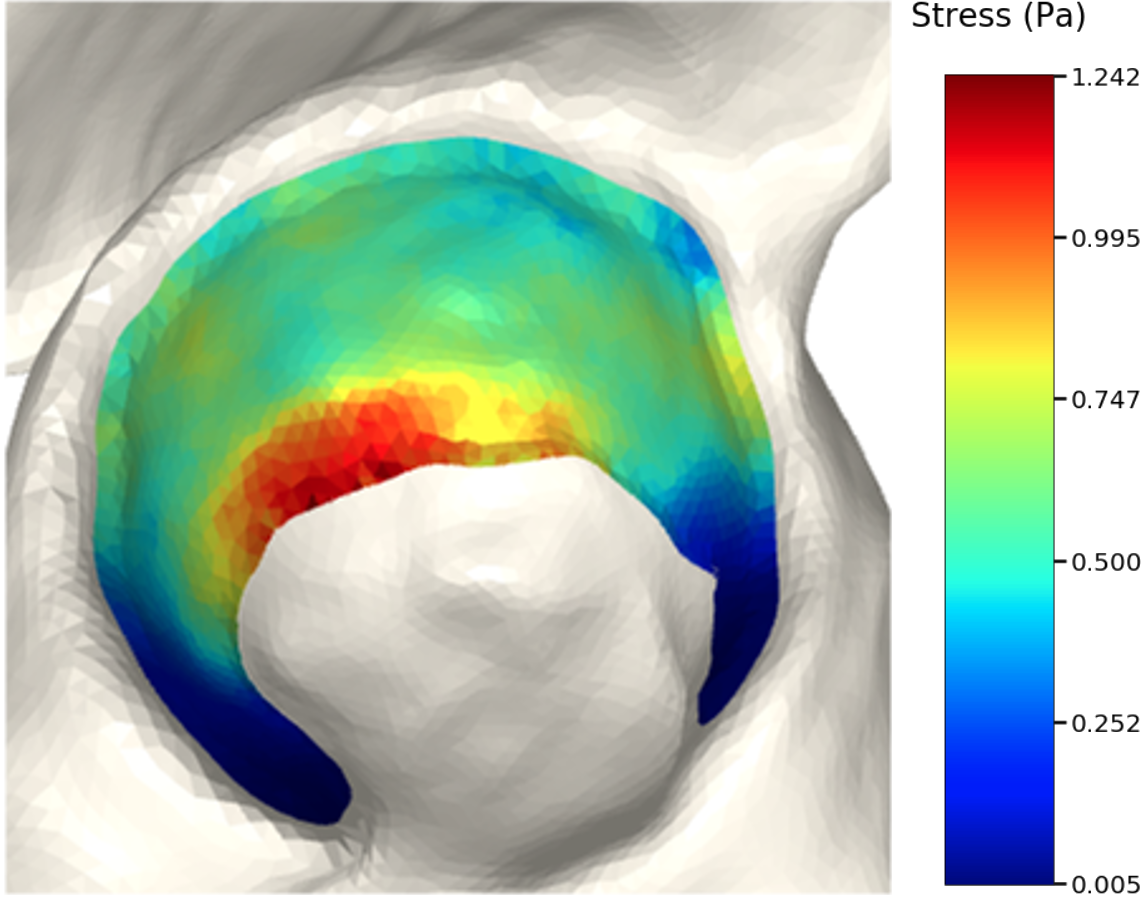}
        \caption{Subject 3 right}
    \end{subfigure}
    \begin{subfigure}[b]{0.3\linewidth}
        \centering
        \includegraphics[width=\linewidth]{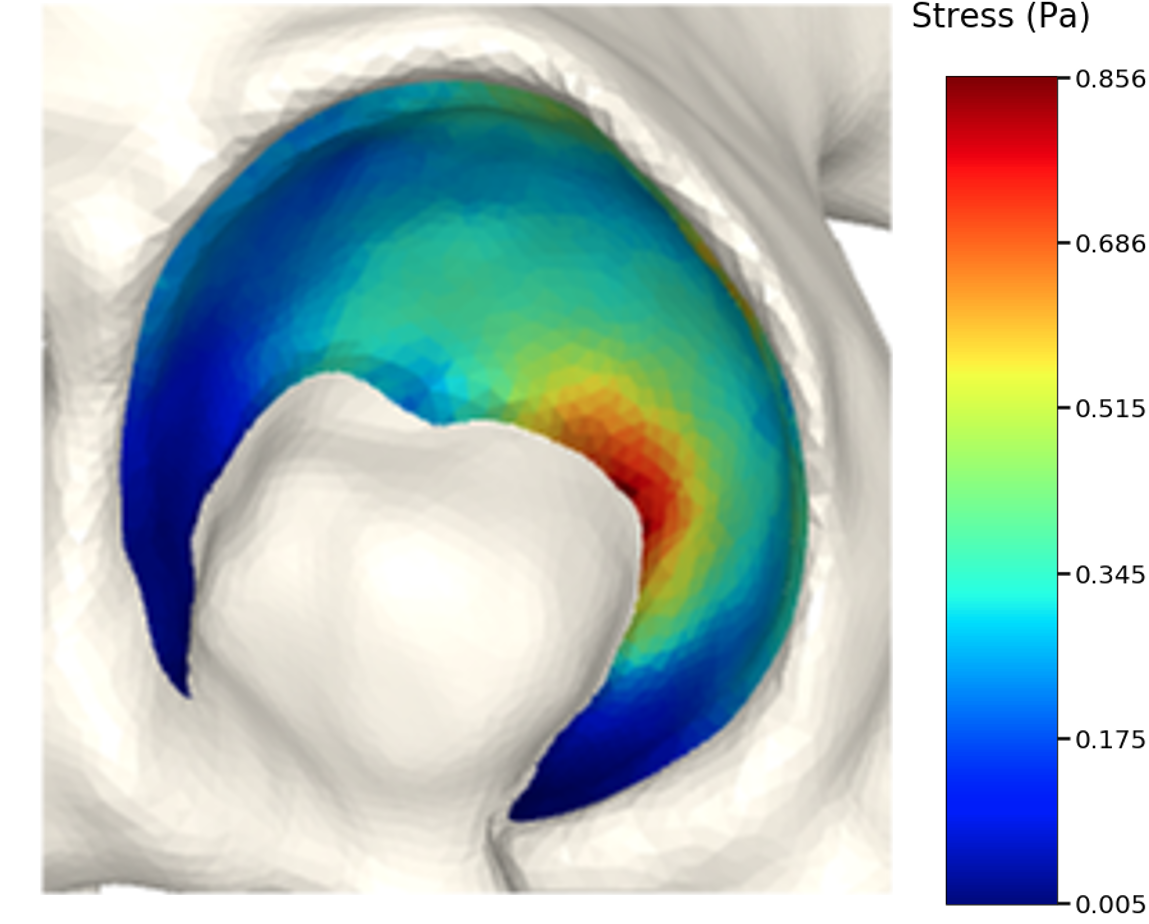}
        \caption{Subject 4 left}
    \end{subfigure}
    \begin{subfigure}[b]{0.3\linewidth}
        \centering
        \includegraphics[width=\linewidth]{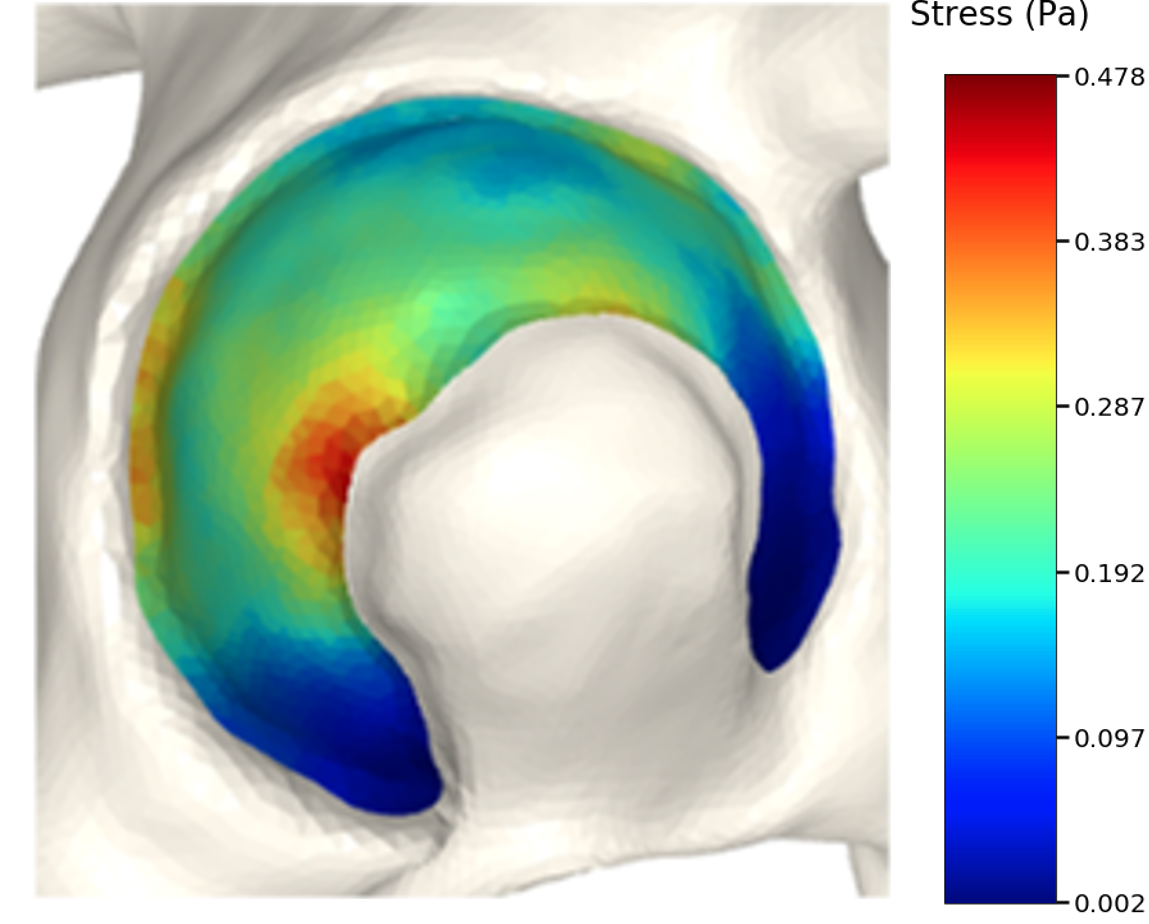}
        \caption{Subject 4 right}
    \end{subfigure}
    \caption{Von Mises stress patterns on the pelvic cartilage of the right side of the subject shown in Fig.~\ref{fig:von:mises:stress} (top left) and the remaining four subjects. We show the left and right pelvic-side cartilage during a simulation identical to the one described in Section~\ref{sec:results} of the main paper for each subject.}
    \label{fig:von:mises:stress:sm}
\end{figure}

\begin{table}[h!]
    
    \centering
    \begin{tabular}{l|
                    >{\centering\arraybackslash}p{0.1\textwidth}|
                    >{\centering\arraybackslash}p{0.1\textwidth}|
                    >{\centering\arraybackslash}p{0.1\textwidth}|
                    >{\centering\arraybackslash}p{0.1\textwidth}|
                    >{\centering\arraybackslash}p{0.1\textwidth}}
         \backslashbox{\textbf{Cartilage}}{\textbf{Parameter}} & 
         $\mathcal{N}$ & $\kappa_{\min}$ & $\kappa_{\max}$ & $\delta$ & $N_{trim}$\\ \hline
         Femoral cartilage & 20 & 0.026 & Inf & 4.2 & 7\\
         Pelvic cartilage & - & - & - & 3 & 2\\
         Sacroiliac  & - & - & - & 4.6 & 0\\
         Pubic & - & - & - & 6 & 1
    \end{tabular}
    \caption{The parameter values used to generate the cartilage from Fig.~\ref{fig:future:joints:hip}.
    The free parameters are the neighbourhood-size used to estimate the curvature of the bone ($\mathcal{N}$); the minimum and maximum curvature in the cartilage region ($\kappa_{\min}, \kappa_{\max}$; Eq.~\eqref{eq:grow}); the distance parameter in $mm$ ($\delta$; Eq.~\ref{eq:distance:filtering}); and the number of times the outer boundary should be trimmed ($N_{trim}$). Here, the curvature based parameters ($\mathcal{N}, \kappa_{\min}, \kappa_{\max})$ are only used for the femur.}
    \label{table:parameters}
\end{table}


\section*{Finite Element Analysis Details}

The bone geometries are discretized using triangle meshes. 
The pelvic and femoral cartilages inherit the same triangle size of their corresponding bones. These values are chosen for the “master” and “slave” domains in the HJ contacting zones to ensure coarser meshes on the pelvis side. We generate 4-node tetrahedral quality volume meshes using the Tetgen package \cite{tetgen}. 
The resulting files are exported as VTK-files to the FEBio solver package \cite{maas2012febio}. FEBio is an open-source non-linear finite element solver which is widely used in biomechanical applications.

We care about observing smooth stress transitions in the cartilage-cartilage interfaces. Hence, for our validation purpose it suffices that the mechanical behavior of all the tissues are set to be homogeneous isotropic linear-elastic material. 
The material properties are based on the review in \cite{R13}. 
The Young’s modulus and Poisson’s ratio is 17GPa and 0.3 for bones and 12MPa and 0.45 for the cartilages.
The bone-cartilage interfaces are modeled as \textit{Tied facet-on-facet} type of contact in FEBio. A \textit{Sliding facet-on-facet} type of contact is applied to the cartilage-cartilage interface.
This is solved as an augmented Lagrangian model with  friction-less tangential interaction allowing unhindered motion in the HJ.

The HJ behavior is analyzed under a quasi-static loading condition. Therefore, the pelvis and the femur is expected to be in an equilibrium state. We use a displacement-controlled simulation where we push the pelvis on top of the femur, representing a pseudo stance position.
We prescribe the nodal displacement field and the load will be given implicitly during the simulation. Considering the stance position, the distal femur is fixed in the x, y, and z-directions. The pelvis is moved in the z-direction
towards the femur.
\end{document}